\documentclass[pmlr,twocolumn,10pt]{jmlr} 
\mlhtrack{proceedings}
\newif\iffinal
\finaltrue  

\iffinal
    \ifmlhneedspmlr
      \jmlrvolume{346}
      \jmlryear{2026}
    \fi
    \ifmlhfindings \jmlrproceedings{}{ML4H 2026 - Findings Track}\fi
    \ifmlhdemo     \jmlrproceedings{}{ML4H 2026 - Demo Track}\fi
    \jmlrworkshop{Machine Learning for Health (ML4H) 2026}
\else
    \jmlrproceedings{}{Submitted to ML4H 2026: \mlhtrackname}
    \jmlrworkshop{Machine Learning for Health (ML4H) 2026}
\fi

\usepackage{graphicx}
\usepackage{placeins}
\usepackage{booktabs}
\usepackage{siunitx}
\usepackage[switch]{lineno}
\usepackage{nicefrac}
\usepackage{enumitem}
\usepackage{stfloats}
\usepackage[most]{tcolorbox}
\usepackage{caption}
\usepackage{tikz}
\usetikzlibrary{arrows.meta,calc}
\usepackage{lato}
\usepackage{multirow}

\theorembodyfont{\upshape}
\theoremheaderfont{\scshape}
\theorempostheader{:}
\theoremsep{\newline}

\title[Interpreting Medical Retrieval with Sparse AutoEncoders]{Explanations that Transfer: Interpreting Medical Retrieval with Sparse AutoEncoders}

\author{%
 \Name{Diane Bouchacourt} \Email{diane.bouchacourt@doctolib.com}\\
 \Name{Ga{\"e}tan Brison} \Email{gaetan.brison@doctolib.com}\\
 \addr Doctolib AI Research Lab, Paris, France
}

\begin{document}

\newcommand{\gaetan}[1]{{\color{blue}#1}}

\maketitle

\ifmlhdemo\else
\begin{abstract}
Medical AI systems that answer users' queries rely on retrieving medical knowledge from large document collections. In this context, fine-tuning general models on medical datasets can improve retrieval relevance. Yet the dense embeddings underlying modern retrieval systems remain largely opaque, making it difficult to understand the mechanics behind their results. Moreover, adapting to a specific medical dataset risks losing transferability to new clinical corpora. In this work, we tackle these issues by applying Sparse Autoencoders (SAEs) to medical information retrieval, transforming dense text embeddings into semantically interpretable representations. To do so, our model combines a reconstruction objective with an InfoNCE contrastive loss that directly preserves query-document relevance. We evaluate three embedding models across six medical retrieval datasets, both in-domain and out-of-domain. Trained on a medical corpus, our sparse codes improve in-domain retrieval over the original embeddings, and lose less out-of-domain performance than their dense counterparts, especially when the target corpus is distant from the training one. Importantly, our experiments show that the method yields interpretable features whose semantic content can be inferred from the queries and documents that activate them most strongly. The interpretation of these features sheds light on why a particular document was retrieved, and transfers beyond the training distribution.
\end{abstract}
\fi

\ifmlhneedsstatements
\paragraph*{Data and Code Availability}
We use 4 \emph{publicly available} benchmarks: CURE \citep{cure2024}, MIRIAD \citep{miriad2025}, NFCorpus \citep{boteva2016nfcorpus}, and R2MED \citep{r2med2025} that  has 3 subsets: Biology, Bioinformatics, and Medical-Sciences, yielding \num{6} datasets in total. Our code is publicly available at \url{https://github.com/doctolib-lab/doctosae}, the weights of the trained SAE models are available at \url{https://huggingface.co/collections/doctolib-lab/doctosae}.

\paragraph*{Institutional Review Board (IRB)}
This work does not require IRB approval: it uses only public benchmark datasets, with no human subjects.
\fi

\section{Introduction}

\begin{figure*}[t]
\floatconts
  {fig:pipeline}
{\caption{
Overview of the proposed pipeline. (1) Medical queries and documents are embedded by (2) a frozen text encoder. (3) A top-k sparse autoencoder (SAE) turns each embedding into a $k$-sparse code, trained with a joint objective (reconstruction and InfoNCE) to match queries to their relevant documents. (4) After training, documents are ranked by cosine similarity using sparse codes. (5) Neurons are interpreted in three ways that can be used to predict activation on texts and explain retrieval dynamics. (6) SAEs and neuron interpretations trained on CURE are evaluated on 5 unseen medical corpora to assess their transferability.}}
{\input{images/sae_pipeline}}
\end{figure*}

Large Language Models (LLMs) have emerged as powerful text encoders that preserve semantic features, and are key in many applications such as information retrieval (IR) and more broadly retrieval-augmented generation where an agent answers a user's question by drawing on external knowledge, such as a database of relevant documents. Typically, a LLM text encoder forms the backbone of the retrieval system, encoding a user's query into an embedding (i.e. a representation vector) which is then matched against a database of documents embeddings to retrieve the most relevant ones. 

In the medical domain, the ability to ground the answer on factual and accurate information is crucial. Effort has been made to develop medical specific text encoders to improve information retrieval \citep{medembed2024}. Still, these models are generally black-box, which impairs understanding of their retrieval dynamics and adoption by health practitioners and users. Some works explore the mechanics behind LLM retrieval \citep{park2025decoding,kang2025interpret} but none specific to medical IR. Yet explainability is key in medical applications in order to ensure safety and accuracy of the generated answer. Moreover, one should ensure the interpretability mechanism is able to generalize to a diversity of datasets, as medical queries are concerned with a variety of topics \citep{boteva2016nfcorpus, cure2024, r2med2025}.

In this work, we bridge this gap and learn an interpretable set of features from dense embeddings via the use of Sparse Autoencoders (SAEs). SAEs have emerged as promising tools for interpreting entangled, high-dimensional representations of large models \citep{ng2011sparse,makhzani2013ksparse} and have later been applied to LLM activations \citep{bricken2023monosemanticity, gao2024scaling, park2025decoding, oneill2024disentangling}. SAEs are trained to project their input into an overcomplete, higher-dimensional basis and reconstruct it as a sparse combination of features in this basis, disentangling complex representations into components that can be interpreted. We train SAEs on the dense embeddings of medical documents and queries, adding a contrastive learning constraint to the SAE original objective, allowing us to preserve IR performance while enabling explainability. Beyond in-domain retrieval performance and interpretability, we study the transferability of the learned features to out-of-domain data. 

\figureref{fig:pipeline} details our pipeline, our contributions are:
\begin{enumerate}[itemsep=0pt,topsep=0.5pt]
\item Section \ref{sec:model}: We train Sparse Autoencoders (SAEs) with a dual reconstruction and retrieval objective to enable interpretability of LLM embedders in medical datasets. We compare three embedders, both generalist and medical, across 6 datasets, both in-domain and out-of-domain.
\item Section \ref{sec:retrievalresults}: We analyze retrieval performance in- and out-of-domain datasets and show that some level of sparsity can help reduce overfitting to the training corpus.
\item Section \ref{sec:interpretation}: We show our learned features are interpretable and generalize to out-of-domain settings, and that retrieval dynamics can be interpreted, as illustrated in \figureref{fig:concept-activation}.
\end{enumerate}
%
%
%
%
%


\definecolor{csRelevant}{HTML}{0B6B6B}   
\definecolor{csIrrelevant}{HTML}{B3261E} 

\newcommand{\csTiny}{\fontsize{5.0}{5.8}\selectfont}       
\newcommand{\csConcept}{\fontsize{5.1}{5.9}\selectfont}    
\newcommand{\csLabel}{\fontsize{6.1}{7.0}\selectfont}      
\newcommand{\csExcerpt}{\fontsize{5.8}{6.8}\selectfont}    
\newcommand{\csNum}{\fontsize{6.8}{7.5}\selectfont}       

\newcommand{\csNote}[1]{%
  \multicolumn{5}{@{}l@{}}{%
    {\csTiny\textit{#1}}%
  }\\[1pt]%
}

\newlength{\csPanel}
\setlength{\csPanel}{0.475\textwidth}

\newlength{\csSpan}
\newlength{\csLabCol}
\newlength{\csValCol}
\newlength{\csQueryBox}

\setlength{\csQueryBox}{2.35em}

\newsavebox{\csBox}

\newcolumntype{C}{>{\centering\arraybackslash}p{\csValCol}}


\newcommand{\csHead}[2]{%
  \parbox[t][18pt][t]{\csValCol}{\centering
    {\csTiny\textcolor{black!55}{#1}\par}%
    {\csConcept #2\par}}}

\newcommand{\csNeurons}[8]{%
  {\csTiny\textsc{neurons}}
    & \csHead{#1}{#2}
    & \csHead{#3}{#4}
    & \csHead{#5}{#6}
    & \csHead{#7}{#8}\\}

%
\newcommand{\csRow}[5]{%
  {\csTiny\textsc{#1}}
    & {\csNum #2}
    & {\csNum #3}
    & {\csNum #4}
    & {\csNum #5}\\}

\newcommand{\csWin}[1]{\textbf{\textcolor{csRelevant}{#1}}}
\newcommand{\csLose}[1]{\textbf{\textcolor{csIrrelevant}{#1}}}


%
\newcommand{\csPassage}[5]{%
  \multicolumn{5}{@{}l@{}}{%
    \sbox{\csBox}{%
      \begin{minipage}{\dimexpr\csSpan-7pt\relax}
        \raggedright
        \setlength{\parskip}{1pt}%
        %
        {\csLabel\textcolor{#1}{\textsc{#2 -- Ranked #3 -- Label: #4}}\par}%
        %
        {\csExcerpt ``#5''\par}%
      \end{minipage}}%
    %
    \textcolor{#1}{%
      \rule[-\dp\csBox]
            {\dimexpr1.5pt\relax}
            {\dimexpr\ht\csBox+\dp\csBox\relax}}%
    \hspace{5pt}\usebox{\csBox}%
  }\\[1pt]}


\newenvironment{cspanel}[1]{%
  \begin{minipage}[t]{\csPanel}%
    \vspace{0pt}%

    \setlength{\csSpan}{\linewidth}%
    \setlength{\csLabCol}{0.130\csSpan}%
    \setlength{\csValCol}{0.2175\csSpan}%
    \setlength{\tabcolsep}{0pt}%

    \parbox[t][\csQueryBox][t]{\csSpan}{%
      {\csTiny\textcolor{black!55}{\textsc{query}}}\quad
      {\footnotesize\emph{#1}}%
    }%

    \par\vspace{2pt}%

    \begin{tabular}{@{}p{\csLabCol}CCCC@{}}
      \toprule
}{%
      \bottomrule
    \end{tabular}%

  \end{minipage}%
}


\begin{figure*}[t]
\centering
\makebox[\textwidth][c]{%


\begin{cspanel}{What is the impact of kidney stones on quality of life?}

  \csNeurons
    {1473}{quality of life}
    {206}{kidney stones}
    {2181}{kidney disease}
    {2677}{pandemic impact}

  \midrule

  \csRow{query}
    {0.581}
    {0.502}
    {0.255}
    {0.226}

  \midrule


  \csPassage
    {csRelevant}
    {Passage A}
    {1}
    {Highly relevant}
    {Urolithiasis negatively impacts quality of life and having more stone episodes worsens quality of life. Certain factors such as chronic medical illnesses and being a primary caregiver for children are associated with lower quality of life.}

  \csNote{\emph{Urolithiasis}: kidney stones.}

  \csRow{passage A}
    {\csWin{0.514}}
    {\csWin{0.486}}
    {0.078}
    {\csWin{0.139}}


  \csPassage
    {csIrrelevant}
    {Passage B}
    {6}
    {Not relevant}
    {Results of the pilot study (67 patients) showed no significant differences in scores depending on whether or not the patients had undergone surgery or had been treated for kidney stones.}

  \csRow{passage B}
    {0.317}
    {0.340}
    {\csLose{0.270}}
    {0.000}

\end{cspanel}

\hspace{0.035\textwidth}%


\begin{cspanel}{What lab monitoring should be done in neonatal patients on
                oral retinoids and at what frequency?}

  \csNeurons
    {1912}{laboratory}
    {2617}{retinoids}
    {2399}{surveillance}
    {1460}{infant health}

  \midrule

  \csRow{query}
    {0.183}
    {0.400}
    {0.220}
    {0.400}

  \midrule


  \csPassage
    {csIrrelevant}
    {Passage A}
    {4}
    {Not relevant}
    {The incidence of isotretinoin-associated laboratory abnormalities varies between different studies, and there is a wide variation in the frequency of laboratory monitoring among different prescribing physicians.}

  \csRow{passage A}
    {\csLose{0.306}}
    {\csLose{0.375}}
    {\csLose{0.229}}
    {0.000}


  \csPassage
    {csRelevant}
    {Passage B}
    {23}
    {Highly relevant}
    {\ldots the retinoid dose should be kept as low as possible, ideally close to 0.5 mg/kg/d. Serum lipids and liver function tests should be performed at baseline, 1 month after initiation of treatment, and every 3 months thereafter. Symptoms suggestive of skeletal toxicity should be promptly investigated, particularly in those children receiving doses greater than 1.0 mg/kg/d.}

  \csRow{passage B}
    {0.000}
    {0.248}
    {0.000}
    {\csWin{0.109}}

\end{cspanel}

}


\caption{\textbf{Sparse concepts explain why one medical retrieval succeeds and another fails.} Two CURE test queries, showing four SAE neurons ($k{=}64$ / BGE embeddings) that most separate the two passages' scores. Stronger activations are bolded in the color of the favored passage. \textcolor{csRelevant}{Highly relevant} is CURE grade~\num{2}; \textcolor{csIrrelevant}{not relevant} is grade~\num{0}. In the right panel, three concepts favor the wrong passage because it states them explicitly, while \emph{infant health}, the population central to the question, favors the passage containing the answer. Excerpts verbatim; full texts in Figure \ref{fig:contrastive-full}. Passage rankings reflect SAE retrieval scores, while highlighted activations reveal the concepts favoring each passage.
}

\label{fig:concept-activation}

\end{figure*}
\vspace{-5mm}
\section{Background}
\paragraph{Sparse Autoencoders interpretations} The superposition hypothesis \citep{elhage2022toy} posits that neural networks, such as Large Language Models (LLMs), represent more features than they have dimensions by encoding them as overlapping directions in their representation space. Sparse Autoencoders (SAEs) \citep{ng2011sparse} aim to resolve this issue by projecting these embeddings onto a higher-dimensional basis where only a few neurons are active. Unlike the dimensions dense embeddings, these sparse features are often substantially more interpretable and closer to single-purpose concepts, making them easier to inspect and analyze \citep{bricken2023monosemanticity,cunningham2023sae}. We use \emph{neuron} and \emph{feature} interchangeably to refer to a single dimension of the sparse code.

SAEs are often applied to internal activations of LLMs, such as those of the residual stream \citep{gao2024scaling} or of the MLP layer \citep{bricken2023monosemanticity}. In both cases, the SAE features are learned on individual token activations in context. \citet{oneill2024disentangling} learn TopK SAEs on embeddings of scientific papers (astronomy and computer science paper abstracts) and show the interpretability of the learned features via an Interpreter-Predictor setting, as well as successful steering of sparse codes. Retrieval performance preservation is not studied. \citet{kang2025interpret} learn SAEs on \gaetan{BAAI General Embedding (BGE)} \citep{xiao2023cpack} embeddings on the MSMARCO \citep{msmarco2016}, with Kullback-Leibler divergence to preserve the query–document relevance structure. The authors interpret their neurons with Neuron-to-Graph \citep{foote2023n2g}, and show that they can manipulate sparse code to steer retrieval. However their quantitative controllability setup is closed-loop: the positive documents are used both to select the features to amplify and to score the resulting gain, showing that relevance is recoverable from the latent space rather than that the right feature can be found. \citet{park2025decoding} train BatchTopK SAEs on SimLM embedding vectors \citep{simlm2023} of MSMARCO, generate descriptions of learned concepts with an LLM, and introduce a BM25-style inverted index over the  neuron that reaches performance competitive with sparse baselines. 

While we train SAEs on dense embeddings, we differ from these works in several ways. First, we focus on medical datasets for both training and evaluation. We preserve query–document relevance through a contrastive InfoNCE loss \citep{oord2018cpc} that leverages ground-truth labels. Importantly, our evaluation studies the transferability of the learned features: we use five out-of-domain datasets and assess the generalization of our method in terms of retrieval performance and explainability. 

\paragraph{Medical Information Retrieval} Several benchmarks evaluate medical retrieval \citep{boteva2016nfcorpus, r2med2025, cure2024}, while general-purpose encoders such as BGE \citep{xiao2023cpack} are adapted to clinical data \citep{medembed2024}. Yet these embedding models lack transparency, preventing understanding of their results.

\raggedbottom
\section{Methodology}
\label{sec:model}

\subsection{SAE on dense embeddings for retrieval}
Figure \ref{fig:pipeline} details our pipeline. A frozen text embedder first maps queries and medical documents to dense embeddings (Steps 1 and 2 in \figureref{fig:pipeline}). Queries embeddings and corpus documents embeddings are denoted $x_q \in \mathbb{R}^d$ and $x_c \in \mathbb{R}^d$ respectively. We then learn the SAE on these embeddings, following the TopK formulation of \citet{gao2024scaling}. In details, given an input embedding $x$ (query or corpus) we compute (Step 3 of \figureref{fig:pipeline}): 
\begin{align}
u &= W_{\mathrm{enc}}(x-b_{\mathrm{pre}})
     + b_{\mathrm{enc}}, \label{eq:preactivation}\\
z &= \mathrm{TopK}_{k}(u), \label{eq:sparse-code}\\
\hat{x} &= W_{\mathrm{dec}}z+b_{\mathrm{pre}}.
\end{align}
where $x \in \mathbb{R}^d, z\in \mathbb{R}^m~\mathrm{with}~m \gg d, \hat{x}\in\mathbb{R}^d$ 
TopK activation selects the $k$ largest neurons and sets all the rest to \num{0}, hence $z$ has at most $k$ non-zero values\footnote{Following \citet{gao2024scaling} we add a ReLU \citep{nair2010rectified} unit after the TopK, to guarantee non-negative activations.}. Following best practices \citep{gao2024scaling, bricken2023monosemanticity} we initialize $b_{\mathrm{pre}}$ in Equation \ref{eq:preactivation} to the geometric median of the training data, initialize the decoder as the transpose of the encoder and normalize the decoder columns to unit norm. 

While \citet{kang2025interpret} preserve relevance by aligning the reconstructed embeddings to the originals with a Kullback-Leibler term, retaining the relevance structure the original BGE embedder already encodes. As they use BGE and the MSMARCO dataset that is a reasonable assumption; we cannot assume the same on medical corpora with the models we use, \texttt{bge-large-en-v1.5} and
\texttt{gemini-embedding-001}, their training data are undocumented. Thus, we leverage the datasets' ground-truth relevance labels via an InfoNCE term on the sparse vectors to improve information retrieval performance on this specific task. We minimize a weighted combination of reconstruction and contrastive losses:
\begin{equation}
\mathcal{L}
=
\mathcal{L}_{\mathrm{rec}}
+
\lambda\mathcal{L}_{\mathrm{NCE}}.
\label{eq:total-loss}
\end{equation}
Once the SAE is trained on the original embeddings, any test query can be fed through the fixed embedding model to obtain $x_q$ and then to the SAE which outputs a $k$-sparse vector $z_q$. Cosine similarity is computed between $z_q$ and the corpus sparse codes to retrieve relevant documents. 

\textbf{Reconstruction loss} A reconstruction term $\mathcal{L}_{\mathrm{rec}}$ encourages the SAE to retain the information contained in the original dense embeddings. At each step we use a batch of medical corpus document embeddings $B_c$ and a batch of query embeddings $B_q$, of sizes $N_c$ and $N_q$ respectively. 
We minimize the reconstruction loss
\begin{equation}
\mathcal{L}_{\mathrm{rec}} = \frac{1}{N_c}\sum_{x_c \in B_c }\lVert \hat{x_c} - x_c \rVert^2 + \frac{1}{N_q}\sum_{x_q \in B_q} \lVert \hat{x_q} - x_q \rVert^2.
\end{equation}

\textbf{InfoNCE constrastive loss} In order to maximally preserve retrieval performance, we add an InfoNCE contrastive term \citep{oord2018cpc} that pulls the sparse vectors of queries closer to their positive documents and pushes away documents that are not relevant. Specifically, at each step we sample a batch of $B$ query–document pairs, where each pair consists of a query $q_i$ and one of its relevant (i.e.\ positive) documents $c_i$, with corresponding sparse vectors (or codes) $z_{q_i}, z_{c_i}$. The InfoNCE term of the batch is defined as:
\begin{equation}
\mathcal{L}_{\mathrm{NCE}}
=
-\frac{1}{B}
\sum_{i=1}^{B}
\log
\frac{
  \exp\!\left(s_{i,i}/\tau\right)
}{
  \exp\!\left(s_{i,i}/\tau\right) +
  \displaystyle
  \sum_{j\in Z_i^-}
  \exp\!\left(s_{i,j}/\tau\right)
}
\label{eq:infonce}
\end{equation}
where $s_{i,j} = \mathrm{sim}(z_{q_i}, z_{c_j})$ denotes the cosine similarity between the query sparse vector $z_{q_i}$ and the document sparse vector $z_{c_j}$, and the set of negatives $Z_i^-$ for query $z_{q_i}$ is composed of the positive corpus documents of the other queries, $z_{c_j},\, j\neq i$ and the batch of documents used for the reconstruction term $\mathcal{L}_{\mathrm{rec}}$. Once trained, we use the features learned by the SAE encoder to (i) perform retrieval via cosine similarity of sparse code (Step 4 of \figureref{fig:pipeline}) and enable interpretation of the neurons' meaning to explain retrieval dynamics (Steps 5 and 6 of \figureref{fig:pipeline}). Details on our training procedure are in Appendix \ref{sec:appdetails}.

\section{Experimental setup}
\label{sec:exp}
We experiment on \num{6} datasets summarized in Table \ref{tab:datasets}. CURE is our training dataset and we also evaluate on out-of-domain (OOD) datasets: we don't train on them; they serve solely to evaluate the transfer of retrieval and interpretability of our method. Datasets corpora can be documents or chunked passages, so throughout the rest of the paper, we use the words corpus ``documents'' or ``passages'' interchangeably. Each query-passage pair carries a ground-truth relevance label. We report Normalized Discounted Cumulative Gain at 10 (NDCG@10, in \%) \citep{jarvelin2002ndcg}, which measures retrieval quality by considering the relevance of the first ten retrieved documents, assigning greater importance to relevant documents appearing higher in the ranking.

\textbf{CURE} \citep{cure2024} is a clinical retrieval benchmark comprising \num{244600} corpus texts and \num{2000} questions spanning ten medical disciplines, ranging from dermatology to pulmonology. We partition the queries into \num{1400} training, \num{200} validation, and \num{400} test. Documents are mostly passages of chunked articles, and relevance labels are highly relevant, moderately relevant, or non-relevant. Our OOD datasets are the following: \textbf{MIRIAD} \citep{miriad2025} draws its corpus from the biomedical literature; we use the evaluation set from
\href{https://huggingface.co/sentence-transformers/embeddinggemma-300m-medical}{EmbeddingGemma-300M-Medical} consisting of \num{1000} test queries and a corpus of \num{31000} passages. Relevance labels are binary (relevant/non-relevant). \textbf{NFCorpus} \citep{boteva2016nfcorpus} is a standard BEIR benchmark \citep{thakur2021beir} consisting of medical nutrition questions. Its corpus has \num{3633} documents and we evaluate on its test set of \num{323} queries, with relevance labels highly relevant, moderately relevant, or non-relevant. \textbf{R2MED} \citep{r2med2025} has three subsets: \textbf{Biology}, \textbf{Bioinformatics}, and \textbf{Medical-Sciences} \num{77} to \num{103} queries and \num{34810} to \num{57359} documents, all with binary relevance labels. 

CURE's corpus is drawn from academic biomedical literature (PubMed Central, Nature, and BioMed Central). NFCorpus is built mostly from PubMed documents, and MIRIAD from full-text medical articles in S2ORC \citep{lo2020s2orc}. Thus, these two are  close to CURE in genre and register. By contrast, the three R2MED subsets draw their queries from StackExchange community Q\&A, with positive documents being the web pages cited in the answers and the remaining corpus drawn from medical Wikipedia; their corpus genre thus differs more from that of CURE. We discuss this in \sectionref{sec:retrievalresults}.

We compare \num{3} embedders: (i) \href{https://huggingface.co/BAAI/bge-large-en-v1.5}{bge-large-en-v1.5} \citep{xiao2023cpack} (BAAI General Embedding, BGE, developed by the Beijing Academy of Artificial Intelligence) (ii) \href{https://huggingface.co/abhinand/MedEmbed-large-v0.1}{MedEmbed-large-v0.1} \citep{medembed2024} (MedEmbed), that is BGE fine-tuned on clinical data and (iii) \href{https://developers.googleblog.com/gemini-embedding-available-gemini-api/}{gemini-embedding-001} \citep{lee2025gemini} (Gemini) a proprietary model exposing only final embeddings. For BGE and MedEmbed $d=1024$ and for Gemini $d=3072$.\looseness=-1

\raggedbottom
\section{Retrieval performance}
\label{sec:retrievalresults}
\begin{table*}[t]
\centering
\small
\floatconts
  {tab:ndcgsparse}
  {\caption{NDCG@10 (\%), mean $\pm$ 95\% CI over seeds. $k$: active SAE latents per embedding. \emph{dense}: $k$ equals the embedding's native dimension. Best $k$ underlined. \textcolor{blue}{Blue}: beats the original embedding.}}
  {%
  \vspace{-15pt}
  \setlength{\tabcolsep}{4pt}
  \setlength{\aboverulesep}{1pt}
  \setlength{\belowrulesep}{1pt}
  \renewcommand{\arraystretch}{0.92}
  \begin{tabular}{@{}c lrrr rrr@{}}
  \toprule
  & & \multicolumn{3}{c}{\emph{academic literature}} & \multicolumn{3}{c}{\emph{StackExchange queries over Wikipedia}} \\
  \cmidrule(lr){3-5}\cmidrule(lr){6-8}
  & &  &  &  & \multicolumn{3}{c}{R2MED} \\
  \cmidrule(lr){6-8}
  & & \multicolumn{1}{c}{CURE} & \multicolumn{1}{c}{NFCorpus} & \multicolumn{1}{c}{MIRIAD} & \multicolumn{1}{c}{Bioinformatics} & \multicolumn{1}{c}{Biology} & \multicolumn{1}{c}{Medical-Sciences} \\
  \midrule
  \multirow{5}{*}{\rotatebox[origin=c]{90}{\textbf{Gemini}}} & original & 52.77 & 42.22 & 84.93 & 51.35 & 32.94 & 52.51 \\
   & k=64 & \textcolor{blue}{58.05{\scriptsize$\,\pm\,$1.33}} & 32.98{\scriptsize$\,\pm\,$3.62} & 76.33{\scriptsize$\,\pm\,$3.79} & 39.84{\scriptsize$\,\pm\,$8.38} & 20.64{\scriptsize$\,\pm\,$8.57} & 39.57{\scriptsize$\,\pm\,$2.60} \\
   & k=128 & \textcolor{blue}{58.76{\scriptsize$\,\pm\,$0.47}} & 35.19{\scriptsize$\,\pm\,$0.23} & 79.07{\scriptsize$\,\pm\,$0.45} & 41.48{\scriptsize$\,\pm\,$2.83} & 23.26{\scriptsize$\,\pm\,$1.30} & 44.23{\scriptsize$\,\pm\,$4.85} \\
   & k=256 & \textcolor{blue}{59.85{\scriptsize$\,\pm\,$0.45}} & 37.42{\scriptsize$\,\pm\,$1.06} & 80.49{\scriptsize$\,\pm\,$0.78} & 44.80{\scriptsize$\,\pm\,$2.66} & \underline{26.36{\scriptsize$\,\pm\,$1.59}} & \underline{47.86{\scriptsize$\,\pm\,$2.50}} \\
   & dense (k=3072) & \textcolor{blue}{\underline{62.07{\scriptsize$\,\pm\,$0.73}}} & \underline{37.73{\scriptsize$\,\pm\,$1.29}} & \underline{84.38{\scriptsize$\,\pm\,$0.68}} & \underline{46.53{\scriptsize$\,\pm\,$2.62}} & 22.65{\scriptsize$\,\pm\,$3.75} & 38.67{\scriptsize$\,\pm\,$5.15} \\
  \addlinespace[2pt]
  \multirow{5}{*}{\rotatebox[origin=c]{90}{\textbf{BGE}}} & original & 53.06 & 38.17 & 77.24 & 27.04 & 12.72 & 27.78 \\
   & k=64 & \textcolor{blue}{57.07{\scriptsize$\,\pm\,$1.25}} & 31.89{\scriptsize$\,\pm\,$0.59} & 68.75{\scriptsize$\,\pm\,$0.49} & 22.50{\scriptsize$\,\pm\,$1.61} & 8.30{\scriptsize$\,\pm\,$2.93} & 22.72{\scriptsize$\,\pm\,$2.93} \\
   & k=128 & \textcolor{blue}{59.35{\scriptsize$\,\pm\,$0.35}} & 34.48{\scriptsize$\,\pm\,$3.70} & 71.83{\scriptsize$\,\pm\,$1.54} & \textcolor{blue}{27.42{\scriptsize$\,\pm\,$3.21}} & 9.76{\scriptsize$\,\pm\,$2.91} & 26.60{\scriptsize$\,\pm\,$3.39} \\
   & k=256 & \textcolor{blue}{60.50{\scriptsize$\,\pm\,$1.02}} & 35.61{\scriptsize$\,\pm\,$0.08} & 73.85{\scriptsize$\,\pm\,$0.92} & \textcolor{blue}{\underline{28.10{\scriptsize$\,\pm\,$0.89}}} & \underline{10.06{\scriptsize$\,\pm\,$1.17}} & \textcolor{blue}{\underline{29.05{\scriptsize$\,\pm\,$0.80}}} \\
   & dense (k=1024) & \textcolor{blue}{\underline{61.87{\scriptsize$\,\pm\,$0.55}}} & \underline{36.62{\scriptsize$\,\pm\,$0.29}} & \underline{75.05{\scriptsize$\,\pm\,$0.36}} & \textcolor{blue}{27.21{\scriptsize$\,\pm\,$1.98}} & 8.37{\scriptsize$\,\pm\,$0.65} & 27.33{\scriptsize$\,\pm\,$1.72} \\
  \addlinespace[2pt]
  \multirow{5}{*}{\rotatebox[origin=c]{90}{\textbf{MedEmbed}}} & original & 52.46 & 37.13 & 76.73 & 25.85 & 8.99 & 25.89 \\
   & k=64 & \textcolor{blue}{56.44{\scriptsize$\,\pm\,$0.84}} & 31.84{\scriptsize$\,\pm\,$0.43} & 69.25{\scriptsize$\,\pm\,$1.17} & 22.51{\scriptsize$\,\pm\,$0.89} & 8.03{\scriptsize$\,\pm\,$2.04} & 22.48{\scriptsize$\,\pm\,$3.16} \\
   & k=128 & \textcolor{blue}{58.49{\scriptsize$\,\pm\,$0.43}} & 34.22{\scriptsize$\,\pm\,$0.84} & 71.76{\scriptsize$\,\pm\,$1.61} & 25.78{\scriptsize$\,\pm\,$0.52} & 8.72{\scriptsize$\,\pm\,$3.18} & \underline{25.32{\scriptsize$\,\pm\,$4.85}} \\
   & k=256 & \textcolor{blue}{59.21{\scriptsize$\,\pm\,$1.08}} & \underline{35.85{\scriptsize$\,\pm\,$0.85}} & 74.44{\scriptsize$\,\pm\,$1.12} & \textcolor{blue}{\underline{28.05{\scriptsize$\,\pm\,$5.68}}} & \underline{8.85{\scriptsize$\,\pm\,$1.81}} & 24.88{\scriptsize$\,\pm\,$3.21} \\
   & dense (k=1024) & \textcolor{blue}{\underline{60.38{\scriptsize$\,\pm\,$0.52}}} & 32.59{\scriptsize$\,\pm\,$0.68} & \underline{74.99{\scriptsize$\,\pm\,$0.31}} & 22.61{\scriptsize$\,\pm\,$0.21} & 6.04{\scriptsize$\,\pm\,$0.97} & 21.99{\scriptsize$\,\pm\,$1.47} \\
  \bottomrule
  \end{tabular}
  }
\end{table*}

\tableref{tab:ndcgsparse} reports NDCG@10 when using cosine similarity over the representations of queries and corpus documents\footnote{With three seeds, $18$ ($3$ embedders x $6$ datasets) uncorrected tests at $\alpha=0.05$ would give a $60\%$ chance of at least one false positive, so we report $95\%$ confidence intervals to show each difference and its uncertainty.}. ``Original'' refers to embeddings produced directly by the pretrained embedding model, without any SAE transformation or additional training. ``Dense'' means $k$ equals the embedders' native dimension $d$ with expansion $m>d$, i.e. a contrastive yet dense model.

\subsection{In-domain: sparsity-retrieval trade-off}
Looking at performance on CURE in \tableref{tab:ndcgsparse}, we see that all $k$ values beat the original baseline performance for all embedders with best performance achieved by the dense models. To the best of our knowledge, BGE and MedEmbed were not trained on CURE's corpus sources, hence why fine-tuning the SAE on the CURE training split helps its test performance. 
Surprisingly, MedEmbed, which is BGE adapted to clinical data, performs slightly worse than BGE on the CURE test queries. As reported in \tableref{tab:ndcgsparse}, the original MedEmbed embeddings achieve an NDCG@10 of $52.46$, compared with $53.06$ for the original BGE embeddings. This difference persists after dense SAE training, where BGE reaches $61.87$ compared with $60.38$ for MedEmbed. Gemini embeddings are the best performing with $62.07$ NDCG@10 with dense ($k=3072$). Regarding the different $k$ values, increasing $k$ exposes the retrieval-sparsity trade-off. For all embedders, $k=128$ and $256$ achieve at least \num{94}\% of the dense performance on CURE.

\tableref{tab:ndcglambda0diff} shows the difference of retrieval performance when fixing $\lambda=0$ (i.e. no InfoNCE term) with respect to \tableref{tab:ndcgsparse}. We see it strongly underperforms in-domain due to the lack of InfoNCE loss. 

\subsection{Out-of-domain: low capacity regularizes supervised domain adaptation}
\tableref{tab:ndcgsparse} shows the three R2MED subsets never benefit from going dense on BGE embeddings, but performance improves with $k$ up to \num{256}.  For BGE, all three subsets are on par with dense at $k=256$, with means above it while $k\le128$ is on par in 5 out of 6 cases. Gemini and MedEmbed behave similarly. Recall that $m$ denotes the total number of dimensions in the SAE latent representation, while $k$ denotes the number of active dimensions per input embedding.

\tableref{tab:allmk} reports the full grid over $k$ and $m$. For dense models in-domain NDCG@10 rises with $m$ while out-of-domain datasets favor small $m$ or don't move. In contrast, sparsity aligns in-domain and out-of-domain performance: for $k\le\num{256}$, the smallest $m$ is better or tied in every dataset. We believe that for dense, the larger capacity $m$ fits the in-domain dataset better at the cost of OOD generalization. The effect of $k$ is mixed on OOD datasets: MIRIAD improves monotonically with $k$ and peaks at dense on all three embedders, as does NFCorpus (except on MedEmbed), in line with these corpora being closer to CURE in genre than R2MED. Thus, for R2MED, dense capacity in most cases buys little over sparse models, and sometimes costs. 

We additionally trained a non-sparse, non-expanded baseline with $k=m=d$ on all three embeddings. In-domain, the contrastive term accounts for most of the gain: on CURE, reconstruction only ($\lambda=0$) the NDCG@10 is 7.5 (BGE) and 8.6 (Gemini) points below the contrastive baseline, and no better than the original embedding. Out-of-domain Table~\ref{tab:nonsparse} shows this baseline performs often comparably to the SAE with $k=256$. Together with the previous finding that expansion helps dense models on CURE but tends to lower performance on the corpora furthest from CURE, this indicates that out-of-domain results are better explained by effective capacity, controlled by both $k$ and $m$. As the SAE align in- and out-of-domain performance with respect to $m$, cross-validating hyper-parameters in-domain make the SAE pick the lowest that generalize better, and the opposite for dense models. Importantly, the next section~\ref{sec:interpretation} shows the SAE's superior interpretability over baselines.

Regarding the constrastive term, $\lambda>0$ gives the best in-domain retrieval but for dense, it is $\lambda=0$ that wins on all OOD datasets (\tableref{tab:ndcglambda0diff}). On the five unseen evaluation datasets, for sparse models with $k \leq \num{256}$, the difference between $\lambda = 0$ and $\lambda > 0$ is small and can favor either configuration. Thus, sparsity reduces the out-of-domain performance degradation when training on CURE.

\section{Interpreting medical features}
\label{sec:interpretation}

Our goal is to provide interpretable features that generalize and enable analysis of retrieval dynamics. As outlined in Step 5 of Figure \ref{fig:pipeline}, our interpretability evaluation proceeds in two phases. First, we generate three types of human-readable interpretations of the neurons learned by the SAEs. Then, we feed these to an Activation Prediction LLM tasked to predict activation of neurons to evaluate these meanings transfer in- and out-of-domain.

\subsection{Phase A: generating interpretations}
\label{sec:neuronsmeanings}

\begin{table}[t]
\centering
\scriptsize
\floatconts
  {tab:neuronexamples}
  {\caption{Example interpretations of a neuron, from the BGE-trained SAE with $k=64$ (only 1 passage and 1 query shown in Snippets, for space). A bracketed [] number is the neuron's activation on the text that follows it. The last row gives its top activating document over Medical-Sciences.}}
  {%
  \vspace{-15pt}
  \setlength{\tabcolsep}{2pt}
  \setlength{\aboverulesep}{0pt}
  \setlength{\belowrulesep}{1pt}
  \renewcommand{\arraystretch}{1}
  \begin{tabular}{@{}l p{0.700\linewidth}@{}}
   & {\small\textbf{Neuron 2177}: bone} \\
  \midrule
  {\small Snippets} & Doc [0.29] A common view on bone tissue engineering portrays four vital aspects for bone regeneration. A scaffold that... \newline Query [0.27] What is the role of cells in these bone grafts? \\
  \addlinespace[3pt]
  {\small TF-IDF terms} & bone, osteomyelitis, cells, bone grafts, recognised \\
  \addlinespace[3pt]
  {\small Interpreter LLM} & bone: This neuron reacts to documents and queries about bone tissue, bone regeneration, bone remodeling, and bone cells. (keywords: bone, regeneration, cells, tissue, remodeling, vascularization) \\
  \addlinespace[3pt]
  \midrule
  {\small Med-Sciences} & [0.27] Osteocyte feedback Feedback from physical activity maintains bone mass, while feedback from osteocytes limits... \\
  \bottomrule
  \end{tabular}
  }
    \vspace{-10pt}
\end{table}

We generate three types of interpretations:

\textbf{(1) Snippets}. These are the \num{5} top activating corpus documents, \num{2} negatives and \num{5} queries previously fed to the Interpreter LLM.

\textbf{(2) TF-IDF interpretations.} Our second way to interpret a neuron is to look at which words appear most often in the texts that activate it, weighted to account for overall vocabulary statistics. We do this with term frequency–inverse document frequency (TF-IDF) \citep{sparckjones1972idf, salton1988termweighting}. Simply said, TF-IDF weights each term in a document by how often it appears in that document times how rare it is across the whole corpus of documents and queries. In our case, TF-IDF is fit once over all training documents and queries, giving a matrix whose rows are those documents and queries and whose columns are the retained unigrams and bigrams. Then for each neuron, we average the TF-IDF rows of its top \num{5} activating corpus documents/passages and \num{5} top activating queries, and take the vocabulary entries with the highest mean weight. 

\textbf{(3) Interpreter LLM output.} For the third type we leverage an Interpreter LLM \citep{oneill2024disentangling, bricken2023monosemanticity}. For each selected neuron, we retrieve its \num{5} highest-activating corpus passages, with additionally \num{2} corpus documents where the neuron did not fire for examples of non-firing documents, and \num{5} highest-activating queries. We provide these examples, together with their activation values (none for the non-firing documents), to an Interpreter LLM \texttt{medgemma-4b-it} using the Interpreter prompt shown in Figure \ref{fig:interpreter-prompt}. The Interpreter produces (i) a short topic label, (ii) a one- to two-sentence description of the concept associated with activation, and (iii) up to ten distinguishing keywords. 
Importantly, each neuron's interpretation (TF-IDF terms, Interpreter LLM output, or snippets) is built once from in-domain data and reused unchanged for every out-of-domain corpus; only the scored text varies across corpora. The Activation Prediction LLM's prompt (shown in Figure \ref{fig:classifier-prompt}) only notes the interpretations were built on another corpus when ran on OOD datasets. Recall that $m$ denotes the total number of dimensions in the SAE latent representation, while $k$ denotes the number of active dimensions per input embedding.

\tableref{tab:neuronexamples} shows interpretations for SAE examplar neuron with $k=64, m=3072$ trained on BGE embeddings. Comparing the three interpretations allows us to assess the generalization of neuron's meaning without relying on a specific interpretation fed to the Activation Prediction LLM.
\subsection{Phase B: scoring}
\label{sec:score-predictor}
Similar to \citet{oneill2024disentangling, bricken2023monosemanticity}, we leverage an Activation Prediction LLM (also \texttt{medgemma-4b-it}) that predicts, from a neuron's interpretation alone, the probability in $[0,1]$ that the neuron fires on a given text. 
If the neurons' meanings are semantically consistent, this probability should align with whether the neuron is firing on that text. We evaluate the three interpretation types. Our evaluation is as follows:
\begin{enumerate}[itemsep=0pt]
    \item Sample up to \num{3} positives and \num{3} negatives evaluation texts per neuron: A text is considered a \emph{positive} for a neuron if that neuron is among its \num{20} strongest active features, and a \emph{negative} if the neuron is inactive. This rank-based definition avoids introducing corpus-specific activation thresholds and can be applied unchanged to out-of-domain datasets.
    We exclude texts used to build interpretations, along with every passage from the same article as those passages. 
    \item Feed the interpretations and texts to the Activation Prediction LLM to output a probability per (neuron, text) pair.
    \item Compute the Pearson correlation $r$ between the Activation Prediction LLM's probability and the ground truth (positive or not).
\end{enumerate}

\tableref{tab:neuronclassif} reports the results for the BGE SAE with $m = 3072$ latent dimensions, of which $k = 64$ are active for each input. The outputs of the Interpreter LLM and TF-IDF terms are on par, with a small advantage for the Interpreter LLM. Snippets are the worst in correlation on every dataset: feeding higher-level concept interpretations to the Activation Prediction LLM works better than simply showing exemplar activating documents/queries. The overall Pearson correlations are positive (random baseline is 0). Recall the interpretations are built solely on the CURE dataset and reused as-is on the OOD datasets: while there is a gap with respect to CURE, the neurons' meanings generalize to a positive correlation, especially for MIRIAD, Medical Sciences and Bioinformatics. Tables \ref{tab:neuronclassif-gemini} and \ref{tab:neuronclassif-medembed} report these numbers for Gemini embeddings ($k=\num{128}$, $m=\num{9216}$) and MedEmbed ($k=\num{64}$, $m=\num{3072}$), sharing similar conclusions. In appendix Table~\ref{tab:neuron_auroc} we report per-neuron AUROC instead of Pearson $r$ for the BGE-trained SAE, and the two correlate between 0.928 and 0.955. 

\begin{table}[t]
\centering
\small
\floatconts
  {tab:neuronclassif}
  {\caption{Activation Prediction Task on neuron's interpretations, from the BGE-trained SAE with $k=64$. $r$ is the Pearson correlation of the predicted probability with the ground truth (positive/negative), averaged over the neurons every interpretation type scored. Per dataset, best $r$ across interpretation types in bold.}}
  {%
  \vspace{-15pt}
  \setlength{\tabcolsep}{2pt}
  \setlength{\aboverulesep}{1pt}
  \setlength{\belowrulesep}{1pt}
  \begin{tabular}{@{}l|r|r|r@{}}
  \toprule
  Dataset & \multicolumn{1}{c|}{TF-IDF terms} & \multicolumn{1}{c|}{Interpreter LLM} & \multicolumn{1}{c}{Snippets} \\
   & $r$ & $r$ & $r$ \\
  \midrule
  CURE & 0.439 & \textbf{0.440} & 0.308 \\
  NFCorpus & 0.236 & \textbf{0.257} & 0.190 \\
  Bioinformatics & 0.276 & \textbf{0.290} & 0.183 \\
  Biology & \textbf{0.235} & 0.226 & 0.142 \\
  Med-Sciences & 0.281 & \textbf{0.286} & 0.173 \\
  MIRIAD & 0.313 & \textbf{0.334} & 0.216 \\
  \bottomrule
  \end{tabular}
  }
\end{table}
These results show that the SAE learns features that can be described in natural language and whose meaning transfers, in part, to corpora the SAE never saw: interpretations built on CURE alone predict activation on all five out-of-domain datasets, for all three embedders. For each embedder and each interpretation type, Activation Prediction performance correlates with retrieval NDCG@10 across the five out-of-domain corpora (Pearson between \num{0.69} and \num{0.94}). In other words, the OOD corpora on which features learned on CURE keep their meaning are also the corpora on which retrieval transfers, suggesting sparse codes transfer when their semantics do. In Table~\ref{tab:interp_baselines} we compare the SAE interpretability results with the original embeddings, and the $k=m=d$ dense, constrastively trained yet not sparse baseline of Section~\ref{sec:retrievalresults}. SAE features are more interpretable than both baselines on every dataset and every description type.

We find the Activation Prediction LLM conservatively places \num{76}\% of its answers at or below \num{0.1}. Yet those values still track the neuron's rank for each text, which is never shown to the LLM: in \tableref{tab:rankband-bge}, \tableref{tab:rankband-gemini} and \tableref{tab:rankband-medembed} the mean probability declines from the strongest band to the weakest in all cases, out of domain included. Additionally, we ran the interpretability evaluation using all ranks \num{1}-{64} in Table~\ref{tab:rankbandbge-full} to show how the mean probability scores is negatively correlated with the rank, meaning the interpretations allow to predict \emph{how strong the neuron will fire} among activated ones. Calibrating LLM predicted probabilities is a known challenge \citep{griot2025metacognition}, which we leave for future work. 

In order to control for the potential biases of MedGemma as Activation Prediction LLM, we also ran the activation prediction using two extra models: Qwen2.5-3BInstruct and Gemma3-4B. The two new Activation prediction models scored the SAE-BGE  trained (k=64, m=3072) and report $r$ in Table~\ref{tab:neuronclassif_llms}. We see all correlations remain positive. Since the three LLms scored exactly the same (neuron, text) pairs we measured their agreement as the Pearson correlation between their predicted scores. Agreement is always positive and is on average highest with the Interpreter LLM descriptions (0.49 to 0.83 across model pairs and datasets), followed by TF-IDF terms (0.42 to 0.71), and lowest with snippets (0.17 to 0.56). Concept-level descriptions (TF-IDF and Interpreter LLM) thus lead different models to consistent predictions, while raw snippets leave more room for model-specific judgments, consistent with snippets scoring lowest in Table~\ref{tab:neuronclassif}.

\paragraph{Human Expert Evaluation.}To further validate the interpretability of our learned features, we conducted a human evaluation involving five medical experts, following the Activation Prediction task described in Appendix~D. We randomly selected 30 neurons from the 60 neurons with the highest interpretability scores, using 15 positive and 15 negative passages. Each expert independently assessed whether a neuron would activate on a given passage, based on its generated description and the corresponding text excerpt. Responses were collected on a four point scale ranging from 1 (Definitely not) to 4 (Definitely yes). Averaging the five experts' predictions yielded a Pearson correlation of $r = 0.81$ with the actual activation labels and an AUROC of $0.92$. The aggregated predictions correctly classified 27 of 30 passages ($90\%$), with three false negatives and no false positives. These results provide encouraging independent human validation of the learned feature interpretations, complementing our automated evaluation.

\subsection{Interpreting retrieval examples}
\label{sec:retrievalexamples}

We now leverage the features learned by our SAEs to shed light on individual retrieval results. \figureref{fig:concept-activation}, using the same SAE as in \tableref{tab:neuronclassif} (BGE embeddings, $k=64$ and $m=3072$), shows two CURE test queries and two retrieved passages, selected from the cosine-similarity ranking over the full corpus.

We print the the four neurons whose activation differs most between Passage A and Passage B. The numbers refer to the neuron's $\ell_2$-normalized activation on each text (query or passage). We ensure the neurons impact the cosine score: these four neurons make up \num{77}\% and \num{76}\% of cosine similarity for each passage respectively in the kidney-stones query, and \num{53}\% and \num{35}\% for the retinoids query. We report the passages' rank by the SAE as well as their CURE ground-truth label (Highly / medium / not relevant), and we provide the neurons' interpretations from Interpreter LLM output as examplar meanings. 

The first query asks about the impact of kidney stones on quality of life. Passage A, whose ground-truth label is highly relevant (grade 2 relevance), is ranked first by the SAE. It reports directly that more stone episodes worsen quality of life; it wins \emph{quality of life}, \emph{kidney stones} and \emph{pandemic impact}. Passage B, ranked sixth and whose ground-truth label is not relevant (grade 0), is a study of a different disease, primary hyperparathyroidism, in which stone patients are a subgroup the authors call too small to yield a result; it wins only \emph{kidney disease}, a broader neuron that any passage naming a kidney condition can trigger. The neurons that detect information relevant to the question give a higher score to the passage that actually answers the question.

The second query asks what lab monitoring neonatal patients on oral retinoids need, and how often. Passage A, ranked fourth, is the right topic in the wrong population: isotretinoin monitoring in acne patients, never mentioning children or neonates, so its \emph{infant health} activation is exactly zero. It wins \emph{laboratory}, \emph{retinoids} and \emph{surveillance}, all words it repeats throughout. Passage B, ranked twenty-third, carries the answer, serum lipids and liver function tests every three months, but buries it in a chunk about survival and writes neither ``laboratory'' nor ``monitoring'', leaving both those neurons at zero. Only \emph{infant health}, the population the question turns on, favors it, and it is outnumbered three to one. Both retrieval illustrations are read the same way, but the second example is the more informative: it names the concepts that outweighed the correct answer. In contrast, the neuron whose meaning relates to the right population, infants, contributes exactly nothing to the passage that won: SAE interpretations can be used to analyze retrieval errors. Complete passages for both are given in \figureref{fig:contrastive-full}. 

\figureref{fig:contrastive-full-ood} displays similar queries-retrieved documents pairs on Medical Sciences and NFCorpus. Section \ref{sec:score-predictor} showed that the neurons' interpretations generalized, here we see the concepts learned in-domain enable understanding of the mechanics behind the retrieval system in a robust manner to out-of-domain. The example in \sectionref{twotopicquery} show the features explain why, on a two-notion query, the SAE promoted relevant passages the original model missed in its top-\num{5}.

\section{Discussion}
In this work, we proposed an interpretable medical retrieval information model that produces explanations while retaining most retrieval performance. On the three R2MED subsets whose corpora differ most from our training corpus, sparse representations match or beat the dense model thanks to capacity reduction, while the dense model’s extra capacity mainly yields in-domain gains. Importantly, the learned features carry meaningful, robust interpretations. Our experiments show interpretation of neurons' meanings, built solely on CURE texts, can be used to predict activations on five unseen medical corpora, with concept-level interpretations (TF-IDF terms or Interpreter LLM output) outperforming raw exemplar snippets. On the OOD datasets, retrieval performance and interpretation transfer move together, suggesting both rest on features that generalize. We then used these interpretations to reveal which concepts activate in queries and documents, thereby providing users with an explanation. 

\paragraph{Limitations}
Our work has several limitations that are interesting future research directions. First, our human clinicians annotation campaign remains small scale, and a larger study could be for future work. Second, our interpretability evaluation relies on an LLM, which may introduce biases or calibration errors, that we try to correct by using several LLMs. Finally, while expected due to capacity constraint, sparse representations do not match dense retrieval performance on all datasets. All models fine-tuned on CURE most often underperform the original embeddings out-of-domain.

\acks{We thank Nicolas Barascud and Bofeng Huang for their insightful comments on the paper, Yijun Zhu and Rapha{ë}l Ginoulhac for their insights, Hermes Martinez and Pierre Dembiermont for compute set-up and Nacim Rahal for project support. We thank the clinicians working with Doctolib for reviewing the interpretations: Raymond Haddad, Thomas Launay, Aurélia Manns, Camille François-Talpaert, Erik Bernard.}

\FloatBarrier
\bibliography{ref}

\newpage
\appendix

\section{Training details}
\label{sec:appdetails}
Our code is fully reproducible up to parallel-computation randomness and batch-size padding affecting LLM outputs. We used AI assistance to write the code. 

\paragraph{Hyperparameter sweep} We sweep over the following grid:
\begin{itemize}[nosep]
\item number of active neurons $k \in \{64, 128, 256\}$, and dense values: $1024$ for BGE/MedEmbed and $3072$ for Gemini (their original embedding sizes);
\item expansion ratio $m/d \in \{3, 9, 21\}$ (we don't experiment with $21$ on Gemini due to computation cost);
\item InfoNCE loss term $\lambda \in \{0, 5\times10^{-6}, 10^{-5}\}$;
\item temperature $\tau \in \{0.01, 0.05, 0.1\}$.
\end{itemize}
The learning rate ($5\times10^{-4}$) and batch size (\num{512}) are held fixed. Each configuration is scored by the mean and standard error of the validation metric over seeds. We select the configuration with the highest average validation NDCG, then, among configurations overlapping in 95\% CIs we take the smallest hidden size $m$, as we found larger $m$ tend to overfit on training corpus MSE. 

\paragraph{Training procedure} Our training processes three independent loaders: a corpus loader and a query loader for the reconstruction (MSE) objective, and a pair loader over positive query-document pairs for the contrastive objective. The corpus is split \num{90}/\num{10} for the reconstruction term only; InfoNCE and validation NDCG@10 always score against the full corpus. The corpus loader, as it is the biggest, defines the epochs, and we cycle over the other two. We train for $100$ epochs and evaluate validation performance every $100$ steps, to keep the best validation NDCG checkpoint. Because the corpus can contain substantially more documents than queries, one epoch passes over queries multiple times for one pass on the training corpus, thus the query contribution in the reconstruction loss is reweighted by $\alpha_q = 0.1$: it prevents the query reconstruction term from being dominated by the corpus term for datasets with a large document-to-query ratio.

Contrastive pairs are batched by a sampler that guarantees each query appears at most once per batch, sampling one positive document per query per batch (uniformly among its positives). The InfoNCE loss takes each query embedding as the anchor and its paired document as the positive; negatives are the other documents in the pairs batch, plus the documents in the MSE-term corpus batch. To ensure that a query doesn't by chance get one of its positives as a negative/distractor, a mask sets the logits of all other known positive documents among its distractors to $-\infty$, so that they disappear from the InfoNCE denominator.

\section{Dataset details}
\label{sec:appdata}
\begin{table*}[h!]
\centering
\small
\setlength{\tabcolsep}{5pt}
\floatconts
  {tab:datasets}
  {\caption{Our benchmark datasets. \emph{Documents} denotes the number of distinct
documents in the corpus; \emph{Queries} refers to the number of queries and
\emph{Rel./query} the number of relevant documents per query. ID~=~in-domain (the SAE is
trained on this corpus), whereas OOD~=~held out. All queries are English.}}
  {%
  \begin{tabular}{@{}llrrrl@{}}
  \toprule
  Dataset & Role & Documents & Queries & Rel./query & Relevance \\
  \midrule
  CURE & ID & 244600 & 400 & 39.3 & graded $\{0,1,2\}$ \\
  \midrule
  MIRIAD & OOD & 31000 & 1000 & 1.0 & binary \\
  NFCorpus & OOD & 3633 & 323 & 38.2 & graded $\{0,1,2\}$ \\
  R2MED Biology & OOD & 57359 & 103 & 3.6 & binary \\
  R2MED Bioinformatics & OOD & 47473 & 77 & 2.9 & binary \\
  R2MED Medical-Sciences & OOD & 34810 & 88 & 2.8 & binary \\
  \bottomrule
  \end{tabular}
  }
\end{table*}

\subsection{MedEmbed fine-tuning corpus vs. CURE corpus overlap}
Because MedEmbed's fine-tuning corpus and CURE are built from similar
sources~\citep{pmc_oa_subset, nature, biomedcentral}, we checked their overlap.
Of \num{27585} MedEmbed passages, exactly one appears verbatim in CURE, and
\num{715} (\SI{2.59}{\percent}) share at least an 8-token span with a CURE
document (longest contiguous match: 11 tokens). The shared spans are generic
clinical boilerplate rather than substantive medical content; the most frequent
are shown in \tableref{tab:medembed-overlap}. The single verbatim passage is:
\begin{quote}\small
``The decision was to make a complete removable overdenture at the maxilla and a
partial removable denture at the mandible. The deciduous teeth were used as
abutments for a complete prosthesis.''
\end{quote}

\begin{table*}[h!]
\small
\floatconts
  {tab:medembed-overlap}
  {\caption{Most frequent 8-token spans shared between MedEmbed fine-tuning
passages and the CURE corpus, with occurrence counts.}}
  {%
  \begin{tabular}{@{}rl@{}}
  \toprule
  Count & Shared span \\
  \midrule
  134 & the patient was admitted to the hospital due \\
   61 & the patient was admitted to the hospital for \\
   57 & patient was advised to follow up with her \\
   24 & stable condition and advised to follow up with \\
   19 & the patient was discharged from the hospital in \\
   17 & year old female patient was admitted to the \\
   17 & discharged in stable condition and advised to follow \\
   14 & the patient was admitted to the intensive care \\
   14 & patient was admitted to the intensive care unit \\
   14 & the patient's postoperative course was uneventful and \\
   12 & to the department of oral and maxillofacial surgery \\
   11 & the patient was discharged from the hospital after \\
   11 & year old male presented to the emergency department \\
   11 & the postoperative course was uneventful and the patient \\
   11 & the patient was transferred to the intensive care \\
  \bottomrule
  \end{tabular}
  }
\end{table*}

\section{Additional retrieval results}
\label{sec:appretrieval}
\begin{figure*}[h!]
\floatconts
  {fig:bge-plot-in}
  {\caption{CURE performance across hyperparameters - BGE embeddings. Retrieval performance (NDCG@10) is shown for different contrastive loss weights $\lambda$ and InfoNCE temperatures $\tau$.}}
  {\centering\includegraphics[width=0.8\textwidth]{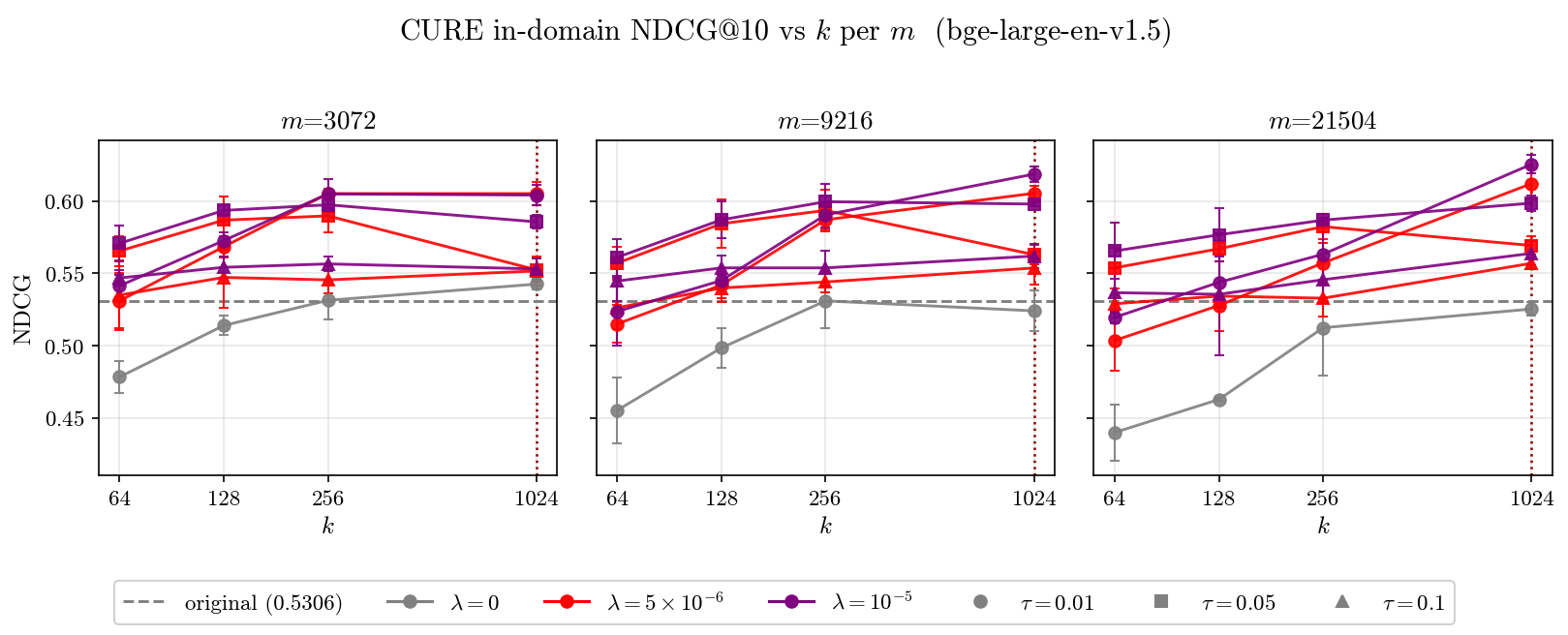}}
\end{figure*}

\begin{figure*}[h!]
\floatconts
  {fig:medembed-plot-in}
  {{\caption{CURE performance across hyperparameters - MedEmbed embedder. Retrieval performance (NDCG@10) is shown for different contrastive loss weights $\lambda$ and InfoNCE temperatures $\tau$.}}}
  {\centering\includegraphics[width=0.8\textwidth]{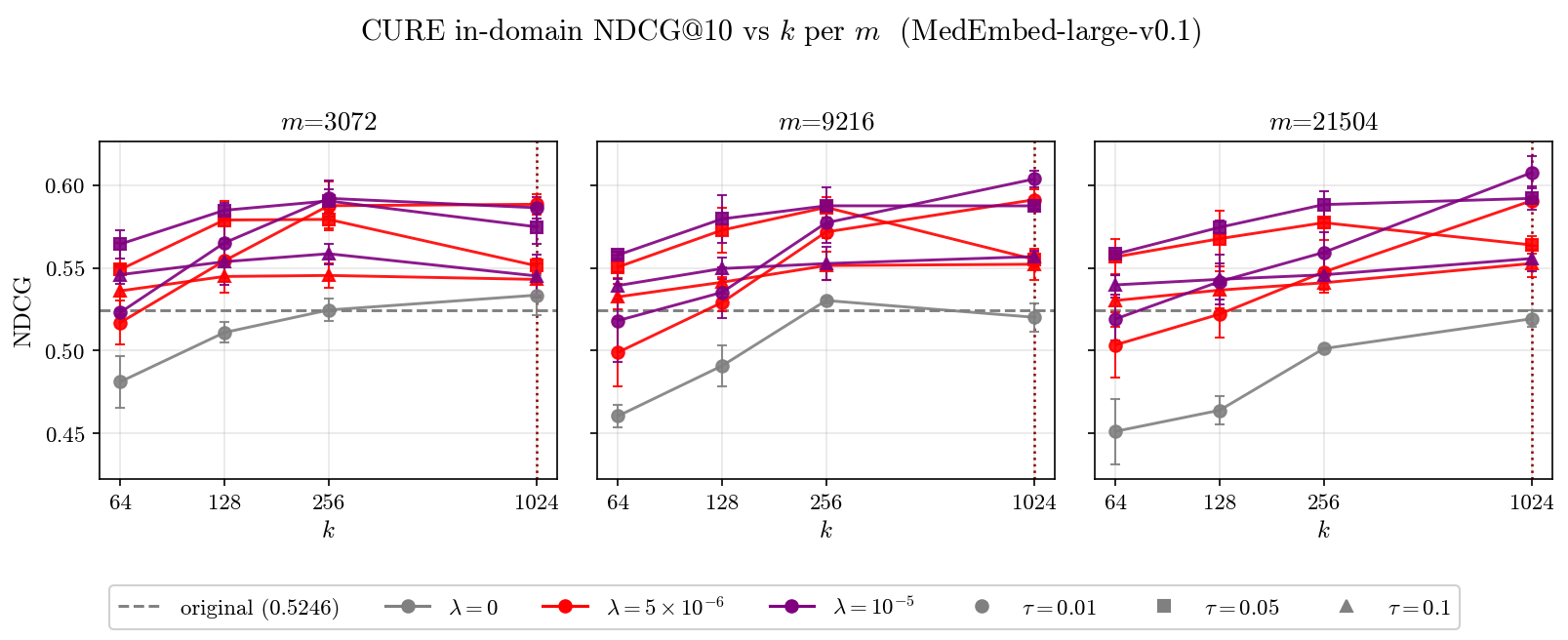}}
\end{figure*}

\begin{figure*}[h!]
\floatconts
  {fig:gemini-plot-in}
  {{\caption{CURE performance across hyperparameters - Gemini embedder. Retrieval performance (NDCG@10) is shown for different contrastive loss weights $\lambda$ and InfoNCE temperatures $\tau$.}}}
  {\centering\includegraphics[width=0.8\textwidth]{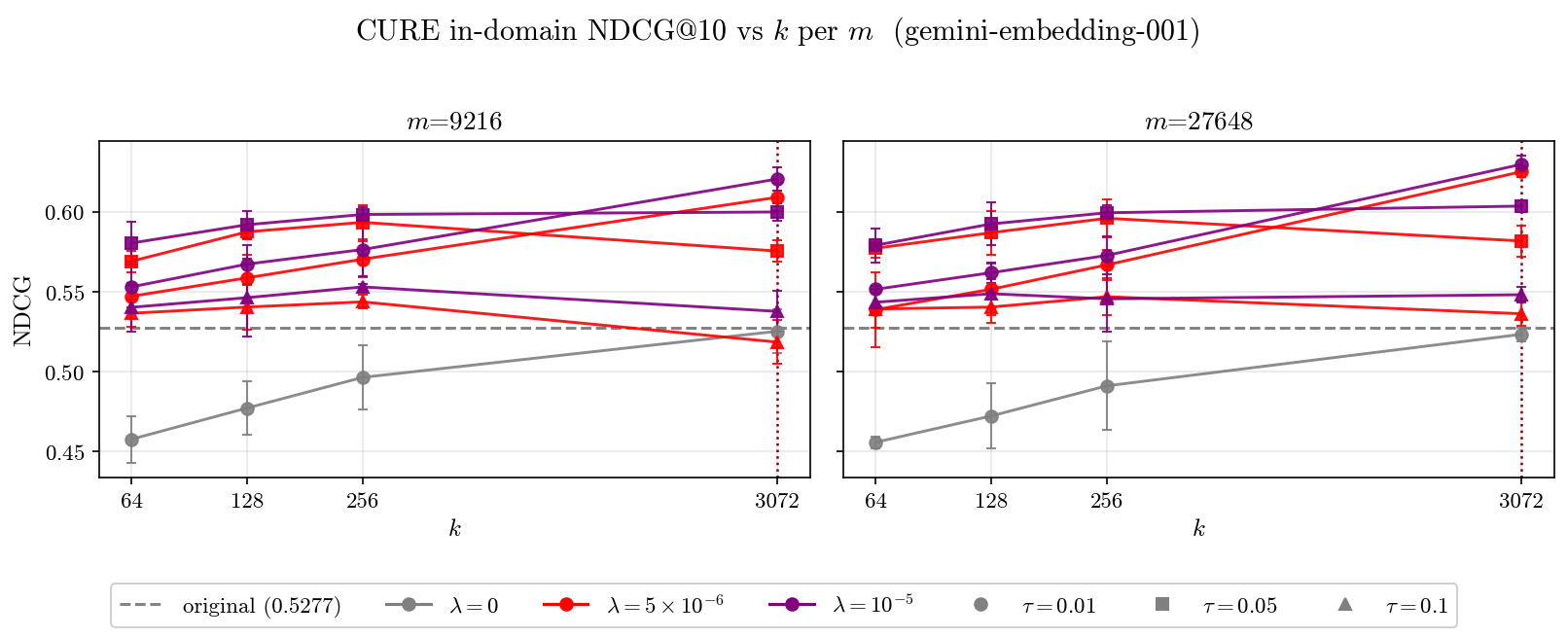}}
\end{figure*}

\begin{figure*}[h!]
\floatconts
  {fig:bge-plot-out-fixed-m}
  {\caption{OOD performance, fixed $m=3072$, comparing the effect of $k$ - BGE embeddings.}}
  {\centering\includegraphics[width=0.8\textwidth]{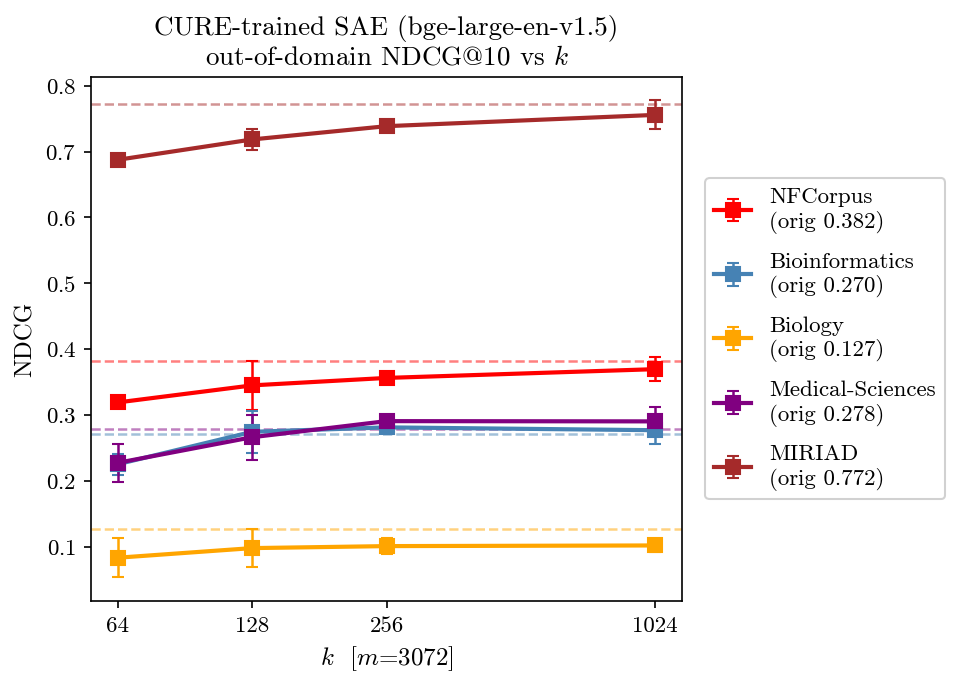}}
\end{figure*}

\begin{figure*}[h!]
\floatconts
  {fig:bge-plot-out-fixed-k}
  {\caption{OOD performance, fixed $k=64$, comparing the effect of $m$ - BGE embeddings.}}
  {\centering\includegraphics[width=0.8\textwidth]{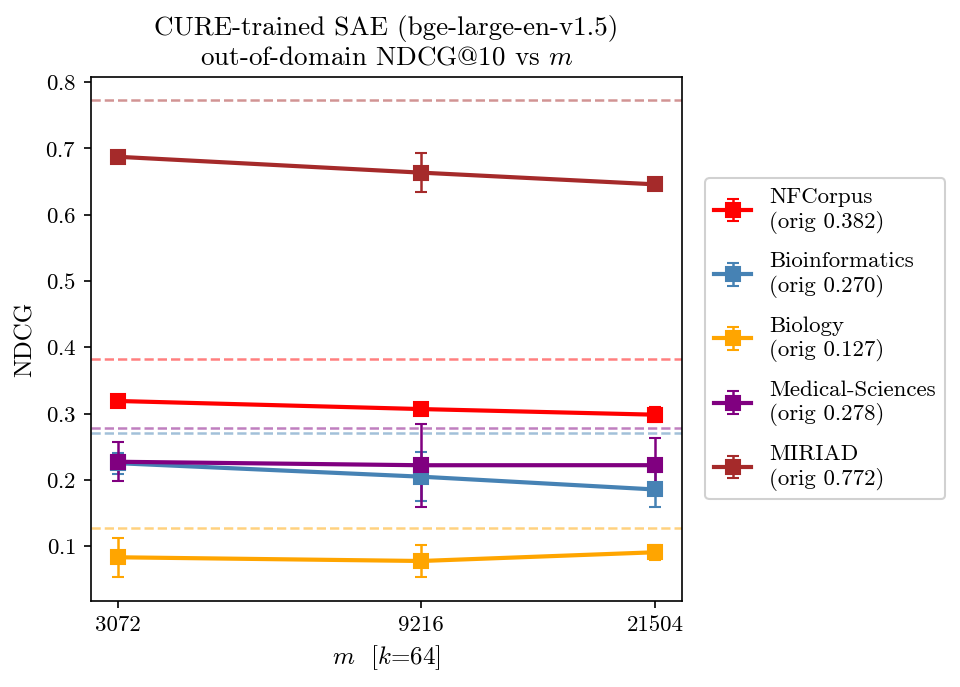}}
\end{figure*}

\begin{figure*}[h!]
\floatconts
  {fig:bge-plot-out-fixed-k2}
  {\caption{OOD performance, fixed $k=64$, comparing the effect of $\lambda$ - BGE embeddings.}}
  {\centering\includegraphics[width=0.8\textwidth]{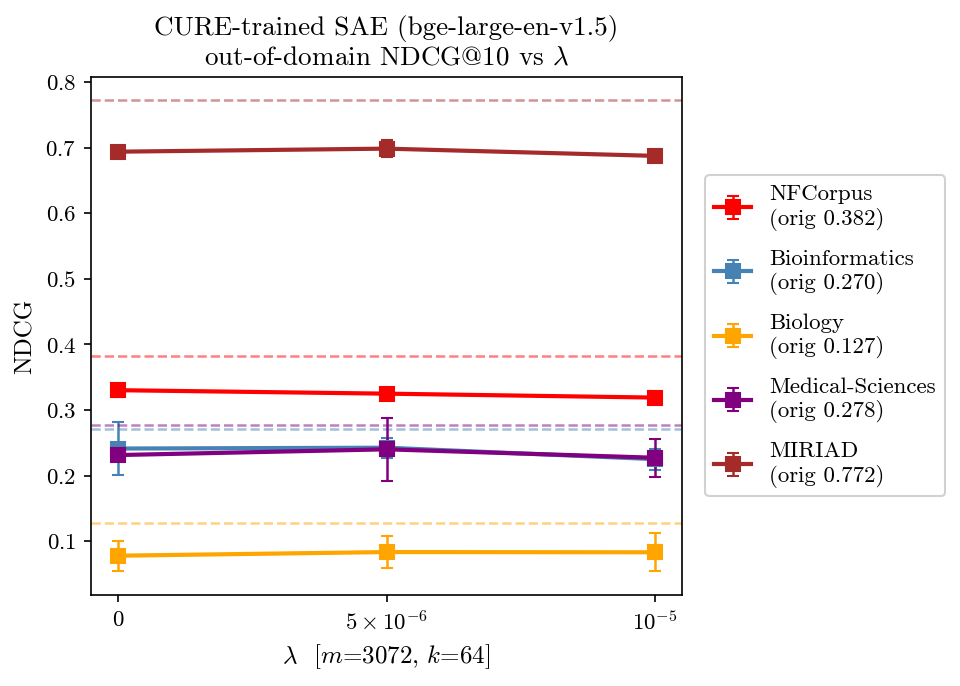}}
\end{figure*}

\begin{table*}[t]
\centering
\small
\floatconts
  {tab:nonsparse}
  {\caption{NDCG@10 (\%), mean $\pm$ 95\% CI over seeds, per dataset. Results are grouped by embedding model (BGE, Gemini, and MedEmbed). SAE: sparse, $k$ active latents ($m=3072$ for BGE and MedEmbed, $m=9216$ for Gemini, selected on validation). \emph{dense}: $k=d$, the embedding's native dimension, with $m>d$. $k{=}m{=}d$: non-sparse and not expanded. Best contrastively trained model per dataset in bold.}}
  {%
  \setlength{\tabcolsep}{4pt}
  \renewcommand{\arraystretch}{1.15}
  \begin{tabular}{@{}lrrrrrr@{}}
  \toprule
  Dataset & $k$=64 & $k$=128 & $k$=256 & dense ($m>d$) & $k{=}m{=}d$ & Original \\
  \midrule
  \multicolumn{7}{l}{\textbf{BGE}} \\
  CURE & 57.07{\scriptsize$\,\pm\,$1.25} & 59.35{\scriptsize$\,\pm\,$0.35} & 60.50{\scriptsize$\,\pm\,$1.02} & \textbf{61.87{\scriptsize$\,\pm\,$0.55}} & 60.18{\scriptsize$\,\pm\,$0.56} & 53.06 \\
  NFCorpus & 31.89{\scriptsize$\,\pm\,$0.59} & 34.48{\scriptsize$\,\pm\,$3.70} & 35.61{\scriptsize$\,\pm\,$0.08} & 36.62{\scriptsize$\,\pm\,$0.29} & \textbf{36.83{\scriptsize$\,\pm\,$2.06}} & 38.17 \\
  MIRIAD & 68.75{\scriptsize$\,\pm\,$0.49} & 71.83{\scriptsize$\,\pm\,$1.54} & 73.85{\scriptsize$\,\pm\,$0.92} & 75.05{\scriptsize$\,\pm\,$0.36} & \textbf{75.27{\scriptsize$\,\pm\,$1.60}} & 77.24 \\
  Bioinformatics & 22.50{\scriptsize$\,\pm\,$1.61} & 27.42{\scriptsize$\,\pm\,$3.21} & \textbf{28.10{\scriptsize$\,\pm\,$0.89}} & 27.21{\scriptsize$\,\pm\,$1.98} & 26.89{\scriptsize$\,\pm\,$2.46} & 27.04 \\
  Biology & 8.30{\scriptsize$\,\pm\,$2.93} & 9.76{\scriptsize$\,\pm\,$2.91} & 10.06{\scriptsize$\,\pm\,$1.17} & 8.37{\scriptsize$\,\pm\,$0.65} & \textbf{10.31{\scriptsize$\,\pm\,$1.15}} & 12.72 \\
  Medical-Sciences & 22.72{\scriptsize$\,\pm\,$2.93} & 26.60{\scriptsize$\,\pm\,$3.39} & \textbf{29.05{\scriptsize$\,\pm\,$0.80}} & 27.33{\scriptsize$\,\pm\,$1.72} & 28.64{\scriptsize$\,\pm\,$0.52} & 27.78 \\
  \midrule
  \multicolumn{7}{l}{\textbf{Gemini}} \\
  CURE & 58.05{\scriptsize$\,\pm\,$1.33} & 58.76{\scriptsize$\,\pm\,$0.47} & 59.85{\scriptsize$\,\pm\,$0.45} & \textbf{62.07{\scriptsize$\,\pm\,$0.73}} & 61.35{\scriptsize$\,\pm\,$0.31} & 52.77 \\
  NFCorpus & 32.98{\scriptsize$\,\pm\,$3.62} & 35.19{\scriptsize$\,\pm\,$0.23} & 37.42{\scriptsize$\,\pm\,$1.06} & \textbf{37.73{\scriptsize$\,\pm\,$1.29}} & 36.73{\scriptsize$\,\pm\,$0.48} & 42.22 \\
  MIRIAD & 76.33{\scriptsize$\,\pm\,$3.79} & 79.07{\scriptsize$\,\pm\,$0.45} & 80.49{\scriptsize$\,\pm\,$0.78} & \textbf{84.38{\scriptsize$\,\pm\,$0.68}} & 83.98{\scriptsize$\,\pm\,$0.42} & 84.93 \\
  Bioinformatics & 39.84{\scriptsize$\,\pm\,$8.38} & 41.48{\scriptsize$\,\pm\,$2.83} & 44.80{\scriptsize$\,\pm\,$2.66} & 46.53{\scriptsize$\,\pm\,$2.62} & \textbf{46.64{\scriptsize$\,\pm\,$1.61}} & 51.35 \\
  Biology & 20.64{\scriptsize$\,\pm\,$8.57} & 23.26{\scriptsize$\,\pm\,$1.30} & \textbf{26.36{\scriptsize$\,\pm\,$1.59}} & 22.65{\scriptsize$\,\pm\,$3.75} & 24.04{\scriptsize$\,\pm\,$1.32} & 32.94 \\
  Medical-Sciences & 39.57{\scriptsize$\,\pm\,$2.60} & 44.23{\scriptsize$\,\pm\,$4.85} & \textbf{47.86{\scriptsize$\,\pm\,$2.50}} & 38.67{\scriptsize$\,\pm\,$5.15} & 38.35{\scriptsize$\,\pm\,$1.63} & 52.51 \\
  \midrule
  \multicolumn{7}{l}{\textbf{MedEmbed}} \\
  CURE & 56.44{\scriptsize$\,\pm\,$0.84} & 58.49{\scriptsize$\,\pm\,$0.43} & 59.21{\scriptsize$\,\pm\,$1.08} & \textbf{60.38{\scriptsize$\,\pm\,$0.52}} & 58.28{\scriptsize$\,\pm\,$0.45} & 52.46 \\
  NFCorpus & 31.84{\scriptsize$\,\pm\,$0.43} & 34.22{\scriptsize$\,\pm\,$0.84} & \textbf{35.85{\scriptsize$\,\pm\,$0.85}} & 32.59{\scriptsize$\,\pm\,$0.68} & 33.07{\scriptsize$\,\pm\,$1.58} & 37.13 \\
  MIRIAD & 69.25{\scriptsize$\,\pm\,$1.17} & 71.76{\scriptsize$\,\pm\,$1.61} & 74.44{\scriptsize$\,\pm\,$1.12} & \textbf{74.99{\scriptsize$\,\pm\,$0.31}} & 74.28{\scriptsize$\,\pm\,$1.96} & 76.73 \\
  Bioinformatics & 22.51{\scriptsize$\,\pm\,$0.89} & 25.78{\scriptsize$\,\pm\,$0.52} & \textbf{28.05{\scriptsize$\,\pm\,$5.68}} & 22.61{\scriptsize$\,\pm\,$0.21} & 23.98{\scriptsize$\,\pm\,$3.32} & 25.85 \\
  Biology & 8.03{\scriptsize$\,\pm\,$2.04} & 8.72{\scriptsize$\,\pm\,$3.18} & \textbf{8.85{\scriptsize$\,\pm\,$1.81}} & 6.04{\scriptsize$\,\pm\,$0.97} & 7.98{\scriptsize$\,\pm\,$1.99} & 8.99 \\
  Medical-Sciences & 22.48{\scriptsize$\,\pm\,$3.16} & \textbf{25.32{\scriptsize$\,\pm\,$4.85}} & 24.88{\scriptsize$\,\pm\,$3.21} & 21.99{\scriptsize$\,\pm\,$1.47} & 23.60{\scriptsize$\,\pm\,$0.74} & 25.89 \\
  \bottomrule
  \end{tabular}
  }
\end{table*}

\begin{table*}[t]
\centering
\floatconts
  {tab:allmk}
  {\caption{NDCG@10 (\%), mean $\pm$ 95\% CI over seeds, for every ($k$, $m$) pair. \emph{dense}: $k$ equals the embedding's native dimension. Remaining hyperparameters are chosen per cell by the best average validation NDCG@10.}}
  {%
  \small
  \vspace{-5pt}
  \setlength{\tabcolsep}{1pt}
  \setlength{\aboverulesep}{1pt}
  \setlength{\belowrulesep}{1pt}
  \begin{tabular}{l|c|cc|ccc}
  \toprule
   & \multicolumn{3}{c|}{\emph{academic literature}} & \multicolumn{3}{c}{\emph{StackExchange queries over Wikipedia}} \\
   & \multicolumn{1}{c|}{} & \multicolumn{2}{c|}{} & \multicolumn{3}{c}{R2MED} \\
   & \multicolumn{1}{c|}{CURE} & \multicolumn{1}{c}{NFCorpus} & \multicolumn{1}{c|}{MIRIAD} & \multicolumn{1}{c}{Bioinformatics} & \multicolumn{1}{c}{Biology} & \multicolumn{1}{c}{Medical-Sciences} \\
  \midrule
  \multicolumn{2}{l|}{\textbf{Gemini}} & \multicolumn{2}{l|}{} & \multicolumn{3}{l}{} \\
  original & 52.77 & 42.22 & 84.93 & 51.35 & 32.94 & 52.51 \\
  \multicolumn{2}{l|}{$k$=64} & \multicolumn{2}{l|}{} & \multicolumn{3}{l}{} \\
  $m$=9216 & 58.05 $\pm$ 1.33 & 32.98 $\pm$ 3.62 & 76.33 $\pm$ 3.79 & 39.84 $\pm$ 8.38 & 20.64 $\pm$ 8.57 & 39.57 $\pm$ 2.60 \\
  $m$=27648 & 57.92 $\pm$ 1.06 & 33.38 $\pm$ 2.73 & 75.74 $\pm$ 0.57 & 38.89 $\pm$ 3.35 & 19.14 $\pm$ 4.22 & 37.47 $\pm$ 2.52 \\
  \multicolumn{2}{l|}{$k$=128} & \multicolumn{2}{l|}{} & \multicolumn{3}{l}{} \\
  $m$=9216 & 58.76 $\pm$ 0.47 & 35.19 $\pm$ 0.23 & 79.07 $\pm$ 0.45 & 41.48 $\pm$ 2.83 & 23.26 $\pm$ 1.30 & 44.23 $\pm$ 4.85 \\
  $m$=27648 & 58.71 $\pm$ 1.38 & 34.77 $\pm$ 0.96 & 77.56 $\pm$ 2.27 & 41.04 $\pm$ 2.65 & 21.38 $\pm$ 2.25 & 43.42 $\pm$ 2.12 \\
  \multicolumn{2}{l|}{$k$=256} & \multicolumn{2}{l|}{} & \multicolumn{3}{l}{} \\
  $m$=9216 & 59.85 $\pm$ 0.45 & 37.42 $\pm$ 1.06 & 80.49 $\pm$ 0.78 & 44.80 $\pm$ 2.66 & 26.36 $\pm$ 1.59 & 47.86 $\pm$ 2.50 \\
  $m$=27648 & 59.61 $\pm$ 1.20 & 36.70 $\pm$ 2.89 & 79.22 $\pm$ 0.60 & 42.00 $\pm$ 1.24 & 24.14 $\pm$ 5.16 & 44.27 $\pm$ 2.75 \\
  \multicolumn{2}{l|}{dense} & \multicolumn{2}{l|}{} & \multicolumn{3}{l}{} \\
  $m$=9216 & 62.07 $\pm$ 0.73 & 37.73 $\pm$ 1.29 & 84.38 $\pm$ 0.68 & 46.53 $\pm$ 2.62 & 22.65 $\pm$ 3.75 & 38.67 $\pm$ 5.15 \\
  $m$=27648 & 62.99 $\pm$ 0.52 & 38.27 $\pm$ 1.61 & 84.15 $\pm$ 0.56 & 47.60 $\pm$ 1.13 & 22.71 $\pm$ 2.15 & 38.45 $\pm$ 2.12 \\
  \midrule
  \multicolumn{2}{l|}{\textbf{BGE}} & \multicolumn{2}{l|}{} & \multicolumn{3}{l}{} \\
  original & 53.06 & 38.17 & 77.24 & 27.04 & 12.72 & 27.78 \\
  \multicolumn{2}{l|}{$k$=64} & \multicolumn{2}{l|}{} & \multicolumn{3}{l}{} \\
  $m$=3072 & 57.07 $\pm$ 1.25 & 31.89 $\pm$ 0.59 & 68.75 $\pm$ 0.49 & 22.50 $\pm$ 1.61 & \phantom{0}8.30 $\pm$ 2.93 & 22.72 $\pm$ 2.93 \\
  $m$=9216 & 56.10 $\pm$ 1.29 & 30.67 $\pm$ 0.69 & 66.35 $\pm$ 2.99 & 20.46 $\pm$ 3.74 & \phantom{0}7.74 $\pm$ 2.41 & 22.19 $\pm$ 6.25 \\
  $m$=21504 & 56.55 $\pm$ 1.94 & 29.82 $\pm$ 1.13 & 64.58 $\pm$ 0.31 & 18.53 $\pm$ 2.68 & \phantom{0}9.06 $\pm$ 1.10 & 22.20 $\pm$ 4.04 \\
  \multicolumn{2}{l|}{$k$=128} & \multicolumn{2}{l|}{} & \multicolumn{3}{l}{} \\
  $m$=3072 & 59.35 $\pm$ 0.35 & 34.48 $\pm$ 3.70 & 71.83 $\pm$ 1.54 & 27.42 $\pm$ 3.21 & \phantom{0}9.76 $\pm$ 2.91 & 26.60 $\pm$ 3.39 \\
  $m$=9216 & 58.70 $\pm$ 1.29 & 31.99 $\pm$ 0.42 & 68.47 $\pm$ 1.29 & 22.22 $\pm$ 4.00 & \phantom{0}9.41 $\pm$ 5.97 & 22.39 $\pm$ 0.86 \\
  $m$=21504 & 57.67 $\pm$ 1.85 & 30.04 $\pm$ 0.76 & 64.47 $\pm$ 2.35 & 19.19 $\pm$ 1.37 & 10.57 $\pm$ 2.33 & 23.30 $\pm$ 4.84 \\
  \multicolumn{2}{l|}{$k$=256} & \multicolumn{2}{l|}{} & \multicolumn{3}{l}{} \\
  $m$=3072 & 60.50 $\pm$ 1.02 & 35.61 $\pm$ 0.08 & 73.85 $\pm$ 0.92 & 28.10 $\pm$ 0.89 & 10.06 $\pm$ 1.17 & 29.05 $\pm$ 0.80 \\
  $m$=9216 & 59.05 $\pm$ 0.87 & 34.92 $\pm$ 1.51 & 71.78 $\pm$ 0.61 & 26.65 $\pm$ 3.85 & 10.21 $\pm$ 2.88 & 28.23 $\pm$ 1.47 \\
  $m$=21504 & 58.68 $\pm$ 0.16 & 32.39 $\pm$ 1.59 & 69.30 $\pm$ 1.10 & 22.75 $\pm$ 6.26 & 11.98 $\pm$ 1.65 & 27.19 $\pm$ 4.29 \\
  \multicolumn{2}{l|}{dense} & \multicolumn{2}{l|}{} & \multicolumn{3}{l}{} \\
  $m$=3072 & 60.42 $\pm$ 0.68 & 36.94 $\pm$ 1.86 & 75.56 $\pm$ 2.19 & 27.67 $\pm$ 2.16 & 10.16 $\pm$ 1.05 & 29.01 $\pm$ 2.20 \\
  $m$=9216 & 61.87 $\pm$ 0.55 & 36.62 $\pm$ 0.29 & 75.05 $\pm$ 0.36 & 27.21 $\pm$ 1.98 & \phantom{0}8.37 $\pm$ 0.65 & 27.33 $\pm$ 1.72 \\
  $m$=21504 & 62.55 $\pm$ 0.61 & 36.73 $\pm$ 0.86 & 75.41 $\pm$ 0.39 & 27.22 $\pm$ 1.29 & \phantom{0}8.61 $\pm$ 0.23 & 27.90 $\pm$ 3.30 \\
  \midrule
  \multicolumn{2}{l|}{\textbf{MedEmbed}} & \multicolumn{2}{l|}{} & \multicolumn{3}{l}{} \\
  original & 52.46 & 37.13 & 76.73 & 25.85 & \phantom{0}8.99 & 25.89 \\
  \multicolumn{2}{l|}{$k$=64} & \multicolumn{2}{l|}{} & \multicolumn{3}{l}{} \\
  $m$=3072 & 56.44 $\pm$ 0.84 & 31.84 $\pm$ 0.43 & 69.25 $\pm$ 1.17 & 22.51 $\pm$ 0.89 & \phantom{0}8.03 $\pm$ 2.04 & 22.48 $\pm$ 3.16 \\
  $m$=9216 & 55.78 $\pm$ 0.21 & 30.47 $\pm$ 1.21 & 67.46 $\pm$ 2.82 & 20.18 $\pm$ 4.09 & \phantom{0}8.19 $\pm$ 2.03 & 22.20 $\pm$ 3.10 \\
  $m$=21504 & 55.84 $\pm$ 0.17 & 29.78 $\pm$ 3.31 & 64.54 $\pm$ 5.91 & 21.62 $\pm$ 1.95 & \phantom{0}8.35 $\pm$ 2.02 & 21.29 $\pm$ 4.13 \\
  \multicolumn{2}{l|}{$k$=128} & \multicolumn{2}{l|}{} & \multicolumn{3}{l}{} \\
  $m$=3072 & 58.49 $\pm$ 0.43 & 34.22 $\pm$ 0.84 & 71.76 $\pm$ 1.61 & 25.78 $\pm$ 0.52 & \phantom{0}8.72 $\pm$ 3.18 & 25.32 $\pm$ 4.85 \\
  $m$=9216 & 57.96 $\pm$ 1.45 & 31.11 $\pm$ 1.52 & 68.51 $\pm$ 0.63 & 21.41 $\pm$ 2.67 & \phantom{0}9.66 $\pm$ 3.50 & 24.17 $\pm$ 4.51 \\
  $m$=21504 & 57.46 $\pm$ 0.44 & 30.22 $\pm$ 1.72 & 65.40 $\pm$ 0.19 & 20.83 $\pm$ 3.73 & \phantom{0}9.00 $\pm$ 2.09 & 23.35 $\pm$ 8.16 \\
  \multicolumn{2}{l|}{$k$=256} & \multicolumn{2}{l|}{} & \multicolumn{3}{l}{} \\
  $m$=3072 & 59.21 $\pm$ 1.08 & 35.85 $\pm$ 0.85 & 74.44 $\pm$ 1.12 & 28.05 $\pm$ 5.68 & \phantom{0}8.85 $\pm$ 1.81 & 24.88 $\pm$ 3.21 \\
  $m$=9216 & 58.66 $\pm$ 0.63 & 34.00 $\pm$ 1.74 & 72.46 $\pm$ 2.17 & 25.99 $\pm$ 2.79 & \phantom{0}9.61 $\pm$ 3.45 & 24.14 $\pm$ 4.94 \\
  $m$=21504 & 58.83 $\pm$ 0.83 & 32.54 $\pm$ 0.78 & 68.46 $\pm$ 3.19 & 21.39 $\pm$ 2.48 & 10.12 $\pm$ 1.77 & 25.38 $\pm$ 0.18 \\
  \multicolumn{2}{l|}{dense} & \multicolumn{2}{l|}{} & \multicolumn{3}{l}{} \\
  $m$=3072 & 58.85 $\pm$ 0.62 & 33.97 $\pm$ 1.95 & 74.65 $\pm$ 1.30 & 24.66 $\pm$ 0.82 & \phantom{0}8.18 $\pm$ 1.17 & 25.12 $\pm$ 0.84 \\
  $m$=9216 & 60.38 $\pm$ 0.52 & 32.59 $\pm$ 0.68 & 74.99 $\pm$ 0.31 & 22.61 $\pm$ 0.21 & \phantom{0}6.04 $\pm$ 0.97 & 21.99 $\pm$ 1.47 \\
  $m$=21504 & 60.77 $\pm$ 0.96 & 33.01 $\pm$ 2.14 & 75.54 $\pm$ 0.78 & 23.23 $\pm$ 2.51 & \phantom{0}6.43 $\pm$ 0.94 & 22.39 $\pm$ 1.35 \\
  \bottomrule
  \end{tabular}
  }
\end{table*}
\begin{table*}[t]
\centering
\small
\floatconts
  {tab:ndcglambda0diff}
  {\caption{NDCG@10 (\%) with $\lambda=0$ minus that of the val-selected $\lambda$ of Table \ref{tab:ndcgsparse}, both means over seeds. Negative: $\lambda=0$ is worse. $k$: active SAE latents; \emph{dense}: $k$ equals the embedding's native dimension.}}
  {%
  \vspace{-15pt}
  \setlength{\tabcolsep}{1pt}
  \setlength{\aboverulesep}{1pt}
  \setlength{\belowrulesep}{1pt}
  \begin{tabular}{l|c|cc|ccc}
  \toprule
   & \multicolumn{3}{c|}{\emph{academic literature}} & \multicolumn{3}{c}{\emph{StackExchange queries over Wikipedia}} \\
   & \multicolumn{1}{c|}{} & \multicolumn{2}{c|}{} & \multicolumn{3}{c}{R2MED} \\
   & \multicolumn{1}{c|}{CURE} & \multicolumn{1}{c}{NFCorpus} & \multicolumn{1}{c|}{MIRIAD} & \multicolumn{1}{c}{Bioinformatics} & \multicolumn{1}{c}{Biology} & \multicolumn{1}{c}{Medical-Sciences} \\
  \midrule
  \multicolumn{7}{l}{\textbf{Gemini}} \\
  k=64 & $-12.29$ & $-0.12$ & $-0.10$ & $+2.03$ & $-0.88$ & $+3.49$ \\
  k=128 & $-11.04$ & $-0.30$ & $+0.23$ & $+3.88$ & $+0.36$ & $+1.62$ \\
  k=256 & $-10.21$ & $+0.52$ & $+0.64$ & $+0.53$ & $+0.48$ & $-0.76$ \\
  dense (k=3072) & $-9.54$ & $+4.83$ & $+0.39$ & $+5.51$ & $+8.59$ & $+14.02$ \\
  \midrule
  \multicolumn{7}{l}{\textbf{BGE}} \\
  k=64 & $-9.25$ & $+1.13$ & $+0.64$ & $+1.62$ & $-0.52$ & $+0.41$ \\
  k=128 & $-7.97$ & $-0.02$ & $+0.86$ & $-1.45$ & $-1.02$ & $+0.22$ \\
  k=256 & $-7.37$ & $+0.98$ & $+2.04$ & $+0.49$ & $+0.18$ & $-1.18$ \\
  dense (k=1024) & $-7.61$ & $+1.17$ & $+1.67$ & $+1.49$ & $+2.77$ & $+1.52$ \\
  \midrule
  \multicolumn{7}{l}{\textbf{MedEmbed}} \\
  k=64 & $-8.32$ & $-0.19$ & $+0.25$ & $+1.77$ & $-1.15$ & $-0.68$ \\
  k=128 & $-7.39$ & $-0.77$ & $+0.72$ & $-1.41$ & $+0.25$ & $-0.66$ \\
  k=256 & $-6.75$ & $+0.36$ & $+1.61$ & $-0.22$ & $+0.36$ & $+3.02$ \\
  dense (k=1024) & $-7.02$ & $+4.56$ & $+1.14$ & $+5.54$ & $+3.42$ & $+6.37$ \\
  \bottomrule
  \end{tabular}
  }
\end{table*}

\begin{table*}[t]
\centering
\small
\floatconts
  {tab:ndcglambda0orig}
  {\caption{NDCG@10 (\%) with $\lambda=0$ minus the original embedding's, means over seeds. Positive: the SAE beats the original. $k$: active SAE latents; \emph{dense}: $k$ equals the embedding's native dimension.}}
  {%
  \vspace{-15pt}
  \setlength{\tabcolsep}{1pt}
  \setlength{\aboverulesep}{1pt}
  \setlength{\belowrulesep}{1pt}
  \begin{tabular}{l|c|cc|ccc}
  \toprule
   & \multicolumn{3}{c|}{\emph{academic literature}} & \multicolumn{3}{c}{\emph{StackExchange queries over Wikipedia}} \\
   & \multicolumn{1}{c|}{} & \multicolumn{2}{c|}{} & \multicolumn{3}{c}{R2MED} \\
   & \multicolumn{1}{c|}{CURE} & \multicolumn{1}{c}{NFCorpus} & \multicolumn{1}{c|}{MIRIAD} & \multicolumn{1}{c}{Bioinformatics} & \multicolumn{1}{c}{Biology} & \multicolumn{1}{c}{Medical-Sciences} \\
  \midrule
  \multicolumn{7}{l}{\textbf{Gemini}} \\
  k=64 & $-7.01$ & $-9.36$ & $-8.70$ & $-9.48$ & $-13.18$ & $-9.45$ \\
  k=128 & $-5.05$ & $-7.33$ & $-5.63$ & $-5.99$ & $-9.31$ & $-6.66$ \\
  k=256 & $-3.13$ & $-4.28$ & $-3.80$ & $-6.01$ & $-6.09$ & $-5.41$ \\
  dense (k=3072) & $-0.24$ & $+0.35$ & $-0.16$ & $+0.69$ & $-1.70$ & $+0.18$ \\
  \midrule
  \multicolumn{7}{l}{\textbf{BGE}} \\
  k=64 & $-5.23$ & $-5.15$ & $-7.86$ & $-2.92$ & $-4.94$ & $-4.65$ \\
  k=128 & $-1.67$ & $-3.71$ & $-4.54$ & $-1.06$ & $-3.98$ & $-0.96$ \\
  k=256 & $+0.07$ & $-1.58$ & $-1.35$ & $+1.55$ & $-2.48$ & $+0.08$ \\
  dense (k=1024) & $+1.19$ & $-0.38$ & $-0.52$ & $+1.66$ & $-1.58$ & $+1.07$ \\
  \midrule
  \multicolumn{7}{l}{\textbf{MedEmbed}} \\
  k=64 & $-4.34$ & $-5.48$ & $-7.24$ & $-1.57$ & $-2.11$ & $-4.10$ \\
  k=128 & $-1.36$ & $-3.68$ & $-4.25$ & $-1.48$ & $-0.02$ & $-1.23$ \\
  k=256 & $-0.00$ & $-0.92$ & $-0.68$ & $+1.98$ & $+0.23$ & $+2.01$ \\
  dense (k=1024) & $+0.90$ & $+0.03$ & $-0.60$ & $+2.30$ & $+0.47$ & $+2.47$ \\
  \bottomrule
  \end{tabular}
  }
\end{table*}

\section{Activation Prediction experiments}
\label{sec:appinterp}

\subsection{Experimental details}
When selecting snippets for interpretation generation or evaluation, CURE documents are chunks of a set of ${\sim}\num{50}$k unique articles, so we capped at 2 documents per unique article. 

When feeding to Interpreter or Activation Prediction LLM we cap each corpus/query at \num{1000} characters.

Because we use neurons firing on $\ge\num{5}$ corpus passages and $\ge\num{5}$ queries for interpretation generation, and select those with at least one positive evaluation text (where OOD enters), the number of scored neurons is smaller than $m$ e.g. for BGE: \num{2379} in-domain, between \num{1779} and \num{2336} out-of-domain.

\subsection{Interpreter and Activation Prediction prompts}
\begin{figure*}[t]
\centering
\small
\floatconts
  {fig:interpreter-prompt}
  {\caption{The Interpreter prompt. Each example body is cut to its first 1000 characters; decoding is greedy, at most 400 new tokens.}}
  {%
  \begin{tcolorbox}[colback=pink!8!white, colframe=magenta!60!black, coltitle=white,
      title={Interpreter Model Prompt}, fonttitle=\bfseries\small, boxrule=1pt,
      arc=1mm, left=3mm, right=3mm, top=2mm, bottom=2mm]
  \textbf{\underline{System message}}
  
  \textbf{Context:} you are an investigator looking at one neuron from a medical AI model. Each neuron fires on certain documents and not on others. Your job is to read the documents it fires on, compare them with documents it does not fire on at all, and say what the first group has in common.
  
  \textbf{Input:} 1 line = 1 document or query, with its activation value in brackets at the start of the line, like "\texttt{[0.28]}". The higher the number, the more strongly the neuron reacted to that one. Where a document has a title, the line reads "\texttt{TITLE:} ... \texttt{|} \texttt{TEXT:} ...", and the neuron saw that whole string, title included. Weight these by their activation value in your prediction. A few example documents that this neuron does not fire on at all are listed at the end, with no activation value shown since it did not fire on them.
  
  Before you answer, work through these steps:
  
  \begin{enumerate}[leftmargin=*, itemsep=1pt, topsep=2pt]
    \item List what the documents and queries it reacts to have in common — topics, themes, methods, wordings — at more than one level of generality, from the broad subject down to the specific.
    \item Use the example documents it does not fire on at all to sharpen your description of what the neuron means: whatever they have in common with the reacting examples is not what makes it react.
    \item Settle on the least complicated account that still fits every reacting example — one theme if one covers them, two if the examples genuinely split, and say so plainly if nothing covers them all.
  \end{enumerate}
  
  \textbf{Output:} fill in these fields.
  
  \begin{itemize}[leftmargin=*, itemsep=1pt, topsep=2pt]
    \item \texttt{"topic"}: the broad medical topic, in 1-3 words (e.g. \texttt{"cardiology"}).
    \item \texttt{"description"}: 1-2 plain sentences on what makes this neuron react, at the level of generality you settled on in step 3 — one theme, or the two it splits into. Don't just repeat the topic, and don't summarise the documents one by one. If the documents share no single topic, say so here.
    \item \texttt{"keywords"}: up to 10 words or short phrases from the documents that best describe this topic, most useful first.
  \end{itemize}
  
  Write everything in plain, simple language. Do not use any abbreviations or acronyms, anywhere, even ones the documents themselves use — spell them out. For example, write "heart attack" instead of "MI", and write out what a program or test does instead of naming it by its letters.
  
  \textbf{Stay accurate:} simplify the words, never the medical facts, and don't add anything that isn't in the documents.
  
  \medskip
  \textbf{\underline{User message}}
  \begin{flushleft}\ttfamily\footnotesize
  Highest~activating~corpus~documents~(5):\\
  ~~[0.31]~TITLE:~<document~title>~|~TEXT:~<document~text>\\
  ~~...~4~more,~by~decreasing~activation\\[4pt]
  Highest~activating~queries~(5):\\
  ~~[0.24]~<query~text>\\
  ~~...~4~more,~by~decreasing~activation\\[4pt]
  A~few~example~documents~that~this~neuron~does~not~fire~on~at~all:\\
  ~~TITLE:~<document~title>~|~TEXT:~<document~text>\\
  ~~TITLE:~<document~title>~|~TEXT:~<document~text>\\[4pt]
  Work~through~the~steps~above,~briefly,~then~close~with~a~JSON~object~and~nothing~after~it:\\
  \{"topic":~"...",~"description":~"...",~"keywords":~["...",~"..."]\}~~//~at~most~10~keywords
  \end{flushleft}
  \end{tcolorbox}
  }
\end{figure*}
\begin{figure*}[t]
\centering
\small
\begin{tcolorbox}[colback=violet!8!white, colframe=violet!65!black, coltitle=white,
    title={(a) Activation Prediction LLM Prompt -- activating snippets}, fonttitle=\bfseries\small, boxrule=1pt,
    arc=1mm, left=3mm, right=3mm, top=2mm, bottom=2mm]
You are a medical expert judging one neuron of a language model trained on medical text. The neuron is described by the examples that activate it most. These are the corpus passages and queries it activates most strongly on, followed by a few example documents it does not fire on at all. Only a small set of the model's neurons is active on any one text: estimate the probability that this neuron is one of them for the text below. Give a number from 0 to 1, where 1 means certain it is active, 0 means certain it is not, and 0.5 means you have no information either way.

One line is one corpus document or one query, opening with that neuron's activation value on it in brackets, e.g. "\texttt{[0.28]} ...". Higher means the neuron fired more strongly on it. A corpus document line reads "\texttt{TITLE:} \texttt{<title>} \texttt{|} \texttt{TEXT:} \texttt{<body>}" where the corpus has titles. The documents listed last, with no activation value, are examples this neuron does not fire on at all. Your answer is a probability, not an activation value: do not copy the scale of the bracketed numbers above.

\textbf{Guidelines:}

\begin{itemize}[leftmargin=*, itemsep=1pt, topsep=2pt]
  \item Match on the specific concept a neuron's examples share, not just the same broad subject. The text may carry topics close to the neuron's without being exactly it; use your judgement.
  \item The neuron's concept can be a secondary theme of the text rather than its main subject.
  \item Do not speculate about topics that are not present in the text. Reason only from what it says.
  \item Weight the examples by their activation values: the strongest ones define what the neuron detects. A low value means it barely fired on that example, so treat it as marginal evidence rather than as a defining case.
\end{itemize}

\medskip
\begin{flushleft}\ttfamily\footnotesize
Neuron~<neuron~id>:\\
<description>\\
<domain~note>\\
Text:\\
<text>\\[4pt]
Respond~with~ONLY~a~JSON~object,~no~other~text:\\
\{"score":~<number~between~0~and~1>\}
\end{flushleft}
\end{tcolorbox}
\end{figure*}

\begin{figure*}[t]
\centering
\small
\begin{tcolorbox}[colback=violet!8!white, colframe=violet!65!black, coltitle=white,
    title={(b) Activation Prediction LLM Prompt -- TF-IDF terms}, fonttitle=\bfseries\small, boxrule=1pt,
    arc=1mm, left=3mm, right=3mm, top=2mm, bottom=2mm]
You are a medical expert judging one neuron of a language model trained on medical text. The neuron is described by its most distinctive terms. These terms are the words that most distinguish the corpus documents it activates most strongly on from the rest of the corpus. Only a small set of the model's neurons is active on any one text: estimate the probability that this neuron is one of them for the text below. Give a number from 0 to 1, where 1 means certain it is active, 0 means certain it is not, and 0.5 means you have no information either way.

\textbf{Guidelines:}

\begin{itemize}[leftmargin=*, itemsep=1pt, topsep=2pt]
  \item Match on the specific concept a neuron's terms point to, not just the same broad subject. The text may carry topics close to the neuron's without being exactly it; use your judgement.
  \item The neuron's concept can be a secondary theme of the text rather than its main subject.
  \item Do not speculate about topics that are not present in the text. Reason only from what it says.
\end{itemize}

\medskip
\begin{flushleft}\ttfamily\footnotesize
Neuron~<neuron~id>:\\
<description>\\
<domain~note>\\
Text:\\
<text>\\[4pt]
Respond~with~ONLY~a~JSON~object,~no~other~text:\\
\{"score":~<number~between~0~and~1>\}
\end{flushleft}
\end{tcolorbox}
\end{figure*}

\begin{figure*}[t]
\centering
\small
\floatconts
  {fig:classifier-prompt}
  {\caption{Activation Prediction LLM prompt, one per neuron interpretation: (a) activating snippets, (b) TF-IDF terms, (c) Interpreter LLM output. \texttt{<domain note>} is empty for an in-domain text, and the following for an out-of-domain one: ``The text below is from a different corpus than the documents this description was built from, so its subject matter, vocabulary and writing style may differ. Judge whether the neuron's own concept is present, not whether the text resembles the corpus the description came from.''}}
  {%
  \begin{tcolorbox}[colback=violet!8!white, colframe=violet!65!black, coltitle=white,
      title={(c) Activation Prediction LLM Prompt -- Interpreter LLM output}, fonttitle=\bfseries\small, boxrule=1pt,
      arc=1mm, left=3mm, right=3mm, top=2mm, bottom=2mm]
  You are a medical expert judging one neuron of a language model trained on medical text. The neuron is described by a sentence naming what it detects. Another model wrote this description after reading the corpus documents it activates most strongly on. Only a small set of the model's neurons is active on any one text: estimate the probability that this neuron is one of them for the text below. Give a number from 0 to 1, where 1 means certain it is active, 0 means certain it is not, and 0.5 means you have no information either way.
  
  \textbf{Guidelines:}
  
  \begin{itemize}[leftmargin=*, itemsep=1pt, topsep=2pt]
    \item Match on the specific concept a neuron's description names, not just the same broad subject. The text may carry topics close to the neuron's without being exactly it; use your judgement.
    \item The neuron's concept can be a secondary theme of the text rather than its main subject.
    \item Do not speculate about topics that are not present in the text. Reason only from what it says.
  \end{itemize}
  
  \medskip
  \begin{flushleft}\ttfamily\footnotesize
  Neuron~<neuron~id>:\\
  <description>\\
  <domain~note>\\
  Text:\\
  <text>\\[4pt]
  Respond~with~ONLY~a~JSON~object,~no~other~text:\\
  \{"score":~<number~between~0~and~1>\}
  \end{flushleft}
  \end{tcolorbox}
  }
\end{figure*}

\subsection{Additional results}
\begin{table}[t]
\centering
\small
\floatconts
  {tab:neuronclassif-gemini}
  {\caption{Activation Prediction Task on neuron's interpretations, from the Gemini-trained SAE with $k=128$. $r$ is the Pearson correlation of the predicted probability with the ground truth (positive/negative), averaged over the neurons every interpretation type scored. Per dataset, best $r$ across interpretation types in bold.}}
  {%
  \setlength{\tabcolsep}{2pt}
  \setlength{\aboverulesep}{1pt}
  \setlength{\belowrulesep}{1pt}
  \begin{tabular}{@{}l|r|r|r@{}}
  \toprule
  Dataset & \multicolumn{1}{c|}{TF-IDF terms} & \multicolumn{1}{c|}{Interpreter LLM} & \multicolumn{1}{c}{Snippets} \\
   & $r$ & $r$ & $r$ \\
  \midrule
  CURE & 0.465 & \textbf{0.491} & 0.360 \\
  NFCorpus & 0.304 & \textbf{0.330} & 0.257 \\
  Bioinformatics & 0.330 & \textbf{0.347} & 0.229 \\
  Biology & 0.246 & \textbf{0.258} & 0.194 \\
  Med-Sciences & 0.319 & \textbf{0.333} & 0.245 \\
  MIRIAD & 0.352 & \textbf{0.385} & 0.261 \\
  \bottomrule
  \end{tabular}
  }
\end{table}

\begin{table}[t]
\centering
\small
\floatconts
  {tab:neuronclassif-medembed}
  {\caption{Activation Prediction Task on neuron's interpretations, from the MedEmbed-trained SAE with $k=64$. $r$ is the Pearson correlation of the predicted probability with the ground truth (positive/negative), averaged over the neurons every interpretation type scored. Per dataset, best $r$ across interpretation types in bold.}}
  {%
  \setlength{\tabcolsep}{2pt}
  \setlength{\aboverulesep}{1pt}
  \setlength{\belowrulesep}{1pt}
  \begin{tabular}{@{}l|r|r|r@{}}
  \toprule
  Dataset & \multicolumn{1}{c|}{TF-IDF terms} & \multicolumn{1}{c|}{Interpreter LLM} & \multicolumn{1}{c}{Snippets} \\
   & $r$ & $r$ & $r$ \\
  \midrule
  CURE & \textbf{0.439} & \textbf{0.439} & 0.320 \\
  NFCorpus & 0.262 & \textbf{0.279} & 0.187 \\
  Bioinformatics & \textbf{0.288} & 0.287 & 0.158 \\
  Biology & 0.203 & \textbf{0.209} & 0.145 \\
  Med-Sciences & 0.278 & \textbf{0.298} & 0.194 \\
  MIRIAD & 0.311 & \textbf{0.359} & 0.241 \\
  \bottomrule
  \end{tabular}
  }
\end{table}

\begin{table}[t]
\centering
\small
\floatconts
  {tab:neuron_auroc}
  {\caption{Activation Prediction Task on neuron's interpretations, from the BGE-trained SAE with $k=64$. AUROC of the predicted probability against the ground truth (positive/negative), averaged over the neurons every interpretation type scored.}}
  {%
  \setlength{\tabcolsep}{2pt}
  \setlength{\aboverulesep}{1pt}
  \setlength{\belowrulesep}{1pt}
  \begin{tabular}{@{}l|rrr@{}}
  \toprule
  Dataset & TF-IDF terms & Interpreter LLM & Snippets \\
  \midrule
  CURE & 0.732 & 0.739 & 0.665 \\
  NFCorpus & 0.625 & 0.635 & 0.600 \\
  Bioinformatics & 0.644 & 0.649 & 0.595 \\
  Biology & 0.623 & 0.620 & 0.577 \\
  Med-Sciences & 0.645 & 0.651 & 0.594 \\
  MIRIAD & 0.665 & 0.677 & 0.619 \\
  \bottomrule
  \end{tabular}
  }
\end{table}

\begin{table}[h]
\centering
\scriptsize
\floatconts
  {tab:rankband-bge}
  {\caption{Mean predicted probability on \emph{positive} evaluation texts, split by the neuron's
  activation rank on the scored text, for the BGE-trained SAE. Every text here is labelled positive
  (the neuron is among the text's \num{20} strongest features); the Activation Predictor is shown neither the rank
  nor any activation value for that text. $\rho$: Spearman between rank and predicted probability, negative because rank \num{1} is the strongest.}}
  {%
  \setlength{\tabcolsep}{4pt}
  \setlength{\aboverulesep}{1pt}
  \setlength{\belowrulesep}{1pt}
  \resizebox{\columnwidth}{!}{%
  \begin{tabular}{@{}ll|rrrr|r@{}}
  \toprule
  Corpus & Interpretation & 1-5 & 6-10 & 11-15 & 16-20 & $\rho$ \\
  \midrule
  CURE      & TF-IDF    & 0.525 & 0.388 & 0.314 & 0.271 & $-0.278$ \\
            & Interpreter LLM & 0.449 & 0.321 & 0.246 & 0.210 & $-0.274$ \\
            & Snippets  & 0.554 & 0.434 & 0.368 & 0.338 & $-0.249$ \\
  \cmidrule{1-7}
  NFCorpus  & TF-IDF    & 0.315 & 0.218 & 0.159 & 0.129 & $-0.211$ \\
            & Interpreter LLM & 0.385 & 0.252 & 0.202 & 0.180 & $-0.217$ \\
            & Snippets  & 0.487 & 0.386 & 0.355 & 0.318 & $-0.175$ \\
  \cmidrule{1-7}
  Bioinf.   & TF-IDF    & 0.345 & 0.244 & 0.186 & 0.153 & $-0.212$ \\
            & Interpreter LLM & 0.363 & 0.247 & 0.190 & 0.163 & $-0.201$ \\
            & Snippets  & 0.448 & 0.369 & 0.343 & 0.307 & $-0.141$ \\
  \cmidrule{1-7}
  Biology   & TF-IDF    & 0.264 & 0.163 & 0.129 & 0.101 & $-0.192$ \\
            & Interpreter LLM & 0.256 & 0.162 & 0.117 & 0.094 & $-0.198$ \\
            & Snippets  & 0.265 & 0.202 & 0.175 & 0.138 & $-0.168$ \\
  \cmidrule{1-7}
  Med-Sci.  & TF-IDF    & 0.334 & 0.204 & 0.161 & 0.138 & $-0.200$ \\
            & Interpreter LLM & 0.344 & 0.221 & 0.165 & 0.135 & $-0.218$ \\
            & Snippets  & 0.392 & 0.301 & 0.260 & 0.241 & $-0.152$ \\
  \cmidrule{1-7}
  MIRIAD    & TF-IDF    & 0.324 & 0.223 & 0.179 & 0.154 & $-0.196$ \\
            & Interpreter LLM & 0.380 & 0.276 & 0.224 & 0.188 & $-0.204$ \\
            & Snippets  & 0.463 & 0.378 & 0.354 & 0.335 & $-0.133$ \\
  \bottomrule
  \end{tabular}
  }
  }
\end{table}

\begin{table}[t]
\centering
\scriptsize
\floatconts
  {tab:rankband-gemini}
  {\caption{As \tableref{tab:rankband-bge}, for the Gemini-trained SAE.}}
  {%
  \setlength{\tabcolsep}{4pt}
  \setlength{\aboverulesep}{1pt}
  \setlength{\belowrulesep}{1pt}
  \resizebox{\columnwidth}{!}{%
  \begin{tabular}{@{}ll|rrrr|r@{}}
  \toprule
  Corpus & Interpretation & 1-5 & 6-10 & 11-15 & 16-20 & $\rho$ \\
  \midrule
  CURE      & TF-IDF    & 0.509 & 0.420 & 0.349 & 0.303 & $-0.222$ \\
            & Interpreter LLM & 0.445 & 0.344 & 0.279 & 0.247 & $-0.211$ \\
            & Snippets  & 0.553 & 0.469 & 0.420 & 0.372 & $-0.205$ \\
  \cmidrule{1-7}
  NFCorpus  & TF-IDF    & 0.342 & 0.228 & 0.201 & 0.160 & $-0.207$ \\
            & Interpreter LLM & 0.414 & 0.278 & 0.250 & 0.217 & $-0.194$ \\
            & Snippets  & 0.492 & 0.392 & 0.364 & 0.322 & $-0.174$ \\
  \cmidrule{1-7}
  Bioinf.   & TF-IDF    & 0.347 & 0.264 & 0.216 & 0.185 & $-0.181$ \\
            & Interpreter LLM & 0.358 & 0.258 & 0.213 & 0.194 & $-0.165$ \\
            & Snippets  & 0.444 & 0.380 & 0.348 & 0.329 & $-0.112$ \\
  \cmidrule{1-7}
  Biology   & TF-IDF    & 0.264 & 0.170 & 0.138 & 0.115 & $-0.173$ \\
            & Interpreter LLM & 0.253 & 0.163 & 0.126 & 0.105 & $-0.180$ \\
            & Snippets  & 0.270 & 0.205 & 0.176 & 0.155 & $-0.139$ \\
  \cmidrule{1-7}
  Med-Sci.  & TF-IDF    & 0.323 & 0.212 & 0.174 & 0.149 & $-0.193$ \\
            & Interpreter LLM & 0.346 & 0.215 & 0.177 & 0.149 & $-0.209$ \\
            & Snippets  & 0.395 & 0.317 & 0.291 & 0.258 & $-0.146$ \\
  \cmidrule{1-7}
  MIRIAD    & TF-IDF    & 0.308 & 0.247 & 0.197 & 0.186 & $-0.157$ \\
            & Interpreter LLM & 0.359 & 0.291 & 0.233 & 0.226 & $-0.156$ \\
            & Snippets  & 0.463 & 0.415 & 0.374 & 0.347 & $-0.139$ \\
  \bottomrule
  \end{tabular}
  }
  }
\end{table}

\begin{table}[t]
\centering
\scriptsize
\floatconts
  {tab:rankband-medembed}
  {\caption{As \tableref{tab:rankband-bge}, for the MedEmbed-trained SAE.}}
  {%
  \setlength{\tabcolsep}{4pt}
  \setlength{\aboverulesep}{1pt}
  \setlength{\belowrulesep}{1pt}
  \begin{tabular}{@{}ll|rrrr|r@{}}
  \toprule
  Corpus & Interpretation & 1-5 & 6-10 & 11-15 & 16-20 & $\rho$ \\
  \midrule
  CURE      & TF-IDF    & 0.518 & 0.396 & 0.309 & 0.262 & $-0.296$ \\
            & Interpreter LLM & 0.436 & 0.308 & 0.244 & 0.197 & $-0.277$ \\
            & Snippets  & 0.537 & 0.419 & 0.357 & 0.319 & $-0.254$ \\
  \cmidrule{1-7}
  NFCorpus  & TF-IDF    & 0.333 & 0.220 & 0.168 & 0.142 & $-0.222$ \\
            & Interpreter LLM & 0.392 & 0.273 & 0.208 & 0.185 & $-0.213$ \\
            & Snippets  & 0.475 & 0.384 & 0.327 & 0.309 & $-0.169$ \\
  \cmidrule{1-7}
  Bioinf.   & TF-IDF    & 0.361 & 0.233 & 0.194 & 0.158 & $-0.204$ \\
            & Interpreter LLM & 0.360 & 0.249 & 0.200 & 0.163 & $-0.200$ \\
            & Snippets  & 0.434 & 0.371 & 0.323 & 0.290 & $-0.156$ \\
  \cmidrule{1-7}
  Biology   & TF-IDF    & 0.254 & 0.154 & 0.119 & 0.107 & $-0.175$ \\
            & Interpreter LLM & 0.240 & 0.158 & 0.101 & 0.094 & $-0.170$ \\
            & Snippets  & 0.263 & 0.184 & 0.163 & 0.145 & $-0.140$ \\
  \cmidrule{1-7}
  Med-Sci.  & TF-IDF    & 0.324 & 0.214 & 0.145 & 0.131 & $-0.222$ \\
            & Interpreter LLM & 0.347 & 0.224 & 0.158 & 0.138 & $-0.216$ \\
            & Snippets  & 0.401 & 0.291 & 0.243 & 0.240 & $-0.161$ \\
  \cmidrule{1-7}
  MIRIAD    & TF-IDF    & 0.322 & 0.244 & 0.187 & 0.160 & $-0.209$ \\
            & Interpreter LLM & 0.387 & 0.284 & 0.236 & 0.196 & $-0.207$ \\
            & Snippets  & 0.459 & 0.391 & 0.358 & 0.322 & $-0.147$ \\
  \bottomrule
  \end{tabular}
  }
\end{table}

\begin{table*}[t]
\centering
\floatconts
  {tab:rankbandbge-full}
  {\caption{As \tableref{tab:rankband-bge}, on all ranks 1-64.}}
  {%
  \setlength{\tabcolsep}{4pt}
  \setlength{\aboverulesep}{1pt}
  \setlength{\belowrulesep}{1pt}
  \begin{tabular}{@{}ll|rrrrrrr|r|r@{}}
  \toprule
  Corpus & Interpretation & 1-5 & 6-10 & 11-15 & 16-20 & 21-30 & 31-45 & 46-64 & Inactive & $\rho$ \\
  \midrule
  CURE      & TF-IDF    & 0.528 & 0.448 & 0.371 & 0.321 & 0.240 & 0.201 & 0.193 & 0.109 & $-0.272$ \\
            & Interpreter LLM & 0.461 & 0.326 & 0.312 & 0.225 & 0.190 & 0.141 & 0.127 & 0.061 & $-0.257$ \\
            & Snippets  & 0.549 & 0.466 & 0.419 & 0.384 & 0.354 & 0.316 & 0.303 & 0.269 & $-0.202$ \\
  \cmidrule{1-11}
  NFCorpus  & TF-IDF    & 0.338 & 0.260 & 0.182 & 0.137 & 0.135 & 0.112 & 0.113 & 0.083 & $-0.133$ \\
            & Interpreter LLM & 0.383 & 0.272 & 0.221 & 0.169 & 0.165 & 0.141 & 0.144 & 0.105 & $-0.109$ \\
            & Snippets  & 0.444 & 0.394 & 0.408 & 0.337 & 0.352 & 0.324 & 0.306 & 0.285 & $-0.094$ \\
  \cmidrule{1-11}
  Bioinf.   & TF-IDF    & 0.356 & 0.301 & 0.232 & 0.189 & 0.147 & 0.133 & 0.119 & 0.075 & $-0.178$ \\
            & Interpreter LLM & 0.366 & 0.294 & 0.217 & 0.181 & 0.147 & 0.132 & 0.125 & 0.074 & $-0.154$ \\
            & Snippets  & 0.461 & 0.399 & 0.353 & 0.318 & 0.316 & 0.300 & 0.294 & 0.254 & $-0.103$ \\
  \cmidrule{1-11}
  Biology   & TF-IDF    & 0.276 & 0.183 & 0.137 & 0.115 & 0.105 & 0.090 & 0.083 & 0.058 & $-0.135$ \\
            & Interpreter LLM & 0.279 & 0.173 & 0.124 & 0.117 & 0.092 & 0.082 & 0.070 & 0.046 & $-0.146$ \\
            & Snippets  & 0.266 & 0.228 & 0.183 & 0.161 & 0.155 & 0.158 & 0.140 & 0.122 & $-0.119$ \\
  \cmidrule{1-11}
  Med-Sci.  & TF-IDF    & 0.323 & 0.227 & 0.188 & 0.137 & 0.124 & 0.116 & 0.091 & 0.056 & $-0.173$ \\
            & Interpreter LLM & 0.335 & 0.257 & 0.187 & 0.148 & 0.128 & 0.111 & 0.098 & 0.057 & $-0.155$ \\
            & Snippets  & 0.388 & 0.340 & 0.291 & 0.264 & 0.267 & 0.243 & 0.230 & 0.191 & $-0.107$ \\
  \cmidrule{1-11}
  MIRIAD    & TF-IDF    & 0.319 & 0.248 & 0.214 & 0.157 & 0.146 & 0.130 & 0.112 & 0.078 & $-0.146$ \\
            & Interpreter LLM & 0.375 & 0.309 & 0.235 & 0.213 & 0.176 & 0.160 & 0.148 & 0.097 & $-0.151$ \\
            & Snippets  & 0.482 & 0.423 & 0.375 & 0.374 & 0.342 & 0.319 & 0.305 & 0.269 & $-0.118$ \\
  \bottomrule
  \end{tabular}
  }
\end{table*}

\begin{table*}[t]
\centering
\floatconts
  {tab:interp_baselines}
  {\caption{Activation prediction results results to gauge interpretability of $k=m=d$ and original embeddings baselines, versus SAE for the BGE embeddings. Pearson correlations between LLM probability and groundtruth (positive or negative).}}
  {%
  \setlength{\tabcolsep}{2pt}
  \setlength{\aboverulesep}{1pt}
  \setlength{\belowrulesep}{1pt}
  \begin{tabular}{@{}l|rrr|rrr|rrr@{}}
  \toprule
   & \multicolumn{3}{c|}{TF-IDF terms} & \multicolumn{3}{c|}{Interpreter LLM} & \multicolumn{3}{c}{Snippets} \\
  Dataset & SAE & $k{=}m{=}d$ & Original & SAE & $k{=}m{=}d$ & Original & SAE & $k{=}m{=}d$ & Original \\
  \midrule
  CURE & \textbf{0.439} & 0.255 & 0.190 & \textbf{0.440} & 0.317 & 0.210 & \textbf{0.308} & 0.184 & 0.141 \\
  NFCorpus & \textbf{0.236} & 0.153 & 0.106 & \textbf{0.257} & 0.191 & 0.149 & \textbf{0.190} & 0.114 & 0.091 \\
  Bioinformatics & \textbf{0.276} & 0.154 & 0.074 & \textbf{0.290} & 0.194 & 0.110 & \textbf{0.183} & 0.115 & 0.056 \\
  Biology & \textbf{0.235} & 0.094 & 0.036 & \textbf{0.226} & 0.138 & 0.052 & \textbf{0.142} & 0.096 & 0.018 \\
  Med-Sciences & \textbf{0.281} & 0.151 & 0.072 & \textbf{0.286} & 0.200 & 0.086 & \textbf{0.173} & 0.114 & 0.052 \\
  MIRIAD & \textbf{0.313} & 0.143 & 0.114 & \textbf{0.334} & 0.205 & 0.202 & \textbf{0.216} & 0.092 & 0.131 \\
  \bottomrule
  \end{tabular}
  }
\end{table*}

\begin{table*}[t]
\centering
\small
\floatconts
  {tab:neuronclassif_llms}
  {\caption{Activation Prediction with three LLMs (Gemma3-4B, MedGemma, Qwen2.5-3B): $r$ is the Pearson correlation of the predicted probability with the ground truth (positive/negative), averaged over the neurons every interpretation type scored.}}
  {%
  \setlength{\tabcolsep}{3pt}
  \setlength{\aboverulesep}{1pt}
  \setlength{\belowrulesep}{1pt}
  \begin{tabular}{@{}lrrrrrr@{}}
  \toprule
  LLM & CURE & NFCorpus & Bioinformatics & Biology & Med-Sciences & MIRIAD \\
  \midrule
  \multicolumn{7}{@{}l}{\textit{TF-IDF terms}} \\
  Gemma3-4B & 0.453 & 0.302 & 0.292 & 0.272 & 0.313 & 0.374 \\
  MedGemma & 0.439 & 0.236 & 0.276 & 0.235 & 0.281 & 0.313 \\
  Qwen2.5-3B & 0.373 & 0.263 & 0.214 & 0.145 & 0.173 & 0.279 \\
  \midrule
  \multicolumn{7}{@{}l}{\textit{Interpreter LLM}} \\
  Gemma3-4B & 0.482 & 0.289 & 0.283 & 0.256 & 0.299 & 0.380 \\
  MedGemma & 0.440 & 0.257 & 0.290 & 0.226 & 0.286 & 0.334 \\
  Qwen2.5-3B & 0.360 & 0.211 & 0.177 & 0.117 & 0.149 & 0.238 \\
  \midrule
  \multicolumn{7}{@{}l}{\textit{Snippets}} \\
  Gemma3-4B & 0.372 & 0.198 & 0.168 & 0.159 & 0.185 & 0.239 \\
  MedGemma & 0.308 & 0.190 & 0.183 & 0.142 & 0.173 & 0.216 \\
  Qwen2.5-3B & 0.382 & 0.146 & 0.111 & 0.041 & 0.083 & 0.147 \\
  \bottomrule
  \end{tabular}%
  }
\end{table*}

\subsection{Qualitative examples}
Figure \ref{fig:contrastive-full} provides the full version of the two
retrieval examples shown in Figure \ref{fig:concept-activation}, including passage
excerpts and the complete feature interpretations.
%
%

\definecolor{cslWon}{HTML}{0B6B6B}
\definecolor{cslLost}{HTML}{A94A2E}
\definecolor{cslGradeOne}{HTML}{B06000}
\definecolor{cslGradeZero}{HTML}{B3261E}
\providecommand{\cspassage}[2]{%
  \sbox0{\begin{minipage}{\dimexpr\linewidth-9pt\relax}#2\end{minipage}}%
  \noindent\textcolor{#1}{\rule[-\dp0]{1.5pt}{\dimexpr\ht0+\dp0\relax}}\hspace{7pt}\usebox0\par}
\begin{figure*}[t]
\begin{minipage}[t]{\linewidth}
\footnotesize
{\scriptsize\bfseries\textsc{CURE}}\\[2pt]
{\scriptsize\textcolor{black!55}{\textsc{query}}}\ \emph{What is the impact of kidney stones on quality of life?}\\[5pt]
\begin{tabular}{@{}lrrrr@{}}
\toprule
{\scriptsize\textsc{neurons}} & \parbox[b]{0.195\linewidth}{\raggedright\scriptsize\textcolor{black!55}{1473}\\\textbf{quality of life}} & \parbox[b]{0.195\linewidth}{\raggedright\scriptsize\textcolor{black!55}{206}\\\textbf{kidney stones}} & \parbox[b]{0.195\linewidth}{\raggedright\scriptsize\textcolor{black!55}{2181}\\\textbf{kidney disease}} & \parbox[b]{0.195\linewidth}{\raggedright\scriptsize\textcolor{black!55}{2677}\\\textbf{pandemic impact}} \\
\midrule
{\scriptsize\textsc{query}} & 0.581 & 0.502 & 0.255 & 0.226 \\
\midrule
\multicolumn{5}{@{}l@{}}{%
\cspassage{cslWon}{%
{\scriptsize\bfseries\textcolor{cslWon}{\textsc{Passage A -- Ranked 1 -- Label: Highly relevant}}}\newline
{\scriptsize\emph{Quality of life and urolithiasis: the patient - reported outcomes measurement information system (PROMIS)}}\newline
{\scriptsize\ Urolithiasis negatively impacts quality of life and having more stone episodes worsens quality of life. Certain factors such as chronic medical illnesses and being a primary caregiver for children are associated with lower quality of life. Medical therapy for stones may improve quality of life and should be considered in those with urolithiasis.}}} \\[3pt]
{\scriptsize\textsc{passage A}} & \textbf{\textcolor{cslWon}{0.514}} & \textbf{\textcolor{cslWon}{0.486}} & 0.078 & \textbf{\textcolor{cslWon}{0.139}} \\
\midrule
\multicolumn{5}{@{}l@{}}{%
\cspassage{cslGradeZero}{%
{\scriptsize\bfseries\textcolor{cslGradeZero}{\textsc{Passage B -- Ranked 6 -- Label: Not relevant}}}\newline
{\scriptsize\emph{Development of a new tool for assessing Health-Related Quality of Life in patients with primary hyperparathyroidism}}\newline
{\scriptsize\ Results of the pilot study (67 patients) showed no significant differences in scores depending on whether or not the patients had undergone surgery or had been treated for kidney stones, but the sample of patients treated with kidney stones was very small. It is to be expected that a larger sample of patients with kidney stones, would show differences between patients with and without symptoms in terms of HRQoL.}}} \\[3pt]
{\scriptsize\textsc{passage B}} & 0.317 & 0.340 & \textbf{\textcolor{cslGradeZero}{0.270}} & 0.000 \\
\bottomrule
\end{tabular}
\end{minipage}\\[14pt]
\begin{minipage}[t]{\linewidth}
\footnotesize
{\scriptsize\bfseries\textsc{CURE}}\\[2pt]
{\scriptsize\textcolor{black!55}{\textsc{query}}}\ \emph{What lab monitoring should be done in neonatal patients on oral retinoids and at what frequency?}\\[5pt]
\begin{tabular}{@{}lrrrr@{}}
\toprule
{\scriptsize\textsc{neurons}} & \parbox[b]{0.195\linewidth}{\raggedright\scriptsize\textcolor{black!55}{1912}\\\textbf{laboratory}} & \parbox[b]{0.195\linewidth}{\raggedright\scriptsize\textcolor{black!55}{2617}\\\textbf{retinoids}} & \parbox[b]{0.195\linewidth}{\raggedright\scriptsize\textcolor{black!55}{2399}\\\textbf{Surveillance}} & \parbox[b]{0.195\linewidth}{\raggedright\scriptsize\textcolor{black!55}{1460}\\\textbf{infant health}} \\
\midrule
{\scriptsize\textsc{query}} & 0.183 & 0.400 & 0.220 & 0.400 \\
\midrule
\multicolumn{5}{@{}l@{}}{%
\cspassage{cslGradeZero}{%
{\scriptsize\bfseries\textcolor{cslGradeZero}{\textsc{Passage A -- Ranked 4 -- Label: Not relevant}}}\newline
{\scriptsize\emph{Laboratory Abnormalities in Acne Patients Treated With Oral Isotretinoin: A Retrospective Epidemiological Study}}\newline
{\scriptsize\ The incidence of isotretinoin-associated laboratory abnormalities varies between different studies, and there is a wide variation in the frequency of laboratory monitoring among different prescribing physicians. The practice of monthly laboratory monitoring among isotretinoin-treated patients lacks evidence and our objective was to address this issue by studying the incidence of these laboratory abnormalities in order to see if frequent monitoring carries a large impact on clinical management on such patients or not. In addition, we studied the effect of body weight and total cumulative dose on laboratory abnormalities.}}} \\[3pt]
{\scriptsize\textsc{passage A}} & \textbf{\textcolor{cslGradeZero}{0.306}} & \textbf{\textcolor{cslGradeZero}{0.375}} & \textbf{\textcolor{cslGradeZero}{0.229}} & 0.000 \\
\midrule
\multicolumn{5}{@{}l@{}}{%
\cspassage{cslWon}{%
{\scriptsize\bfseries\textcolor{cslWon}{\textsc{Passage B -- Ranked 23 -- Label: Highly relevant}}}\newline
{\scriptsize\emph{Harlequin Ichthyosis A Review of Clinical and Molecular Findings in 45 Cases}}\newline
{\scriptsize\ A previous study14 showed that 6 of 7 cases of HI treated with an oral retinoid survived beyond 1 year. Our results demonstrate that the majority of survivors were treated with a retinoid and 80\% started treatment by day 7. Liquid acitretin is not widely available and may take several days to obtain. We postulate that the introduction of oral retinoids, as soon as possible after birth, probably improves survival. However, we recognize that a more active management approach overall also contributes to improved survival. The only potential retinoid related adverse effect in this study was the development of a cataract in a child treated with acitretin (0.5 mg/kg/d) for 4 months from birth. Retinoid-associated cataracts have been reported in adults but not in children.1 To minimize long-term toxic effects, Brecher et al1 recommended that the retinoid dose should be kept as low as possible, ideally close to 0.5 mg/kg/d. Serum lipids and liver function tests should be performed at baseline, 1 month after initiation of treatment, and every 3 months thereafter. Symptoms suggestive of skeletal toxicity should be promptly investigated, particularly in those children receiving doses greater than 1.0 mg/kg/d.}}} \\[3pt]
{\scriptsize\textsc{passage B}} & 0.000 & 0.248 & 0.000 & \textbf{\textcolor{cslWon}{0.109}} \\
\bottomrule
\end{tabular}
\end{minipage}
\caption{\textbf{The two retrieval examples of Figure~\ref{fig:concept-activation}, with complete passages.} Same SAE (BGE, $k{=}64$, $m{=}3072$) and four neurons per panel whose contributions differ most between passages. Each numeric row gives one activation per neuron, in the column order of the header above it. Stronger firing is bolded in the color of the passage it favors. Passages are rank-ordered over \num{244600} CURE passages. \textcolor{cslWon}{Highly relevant} is CURE grade~\num{2}; \textcolor{cslGradeZero}{not relevant} is grade~\num{0}.}
\label{fig:contrastive-full}
\end{figure*}

%
%

\definecolor{cslWon}{HTML}{0B6B6B}
\definecolor{cslLost}{HTML}{A94A2E}
\definecolor{cslGradeOne}{HTML}{B06000}
\definecolor{cslGradeZero}{HTML}{B3261E}
\providecommand{\cspassage}[2]{%
  \sbox0{\begin{minipage}{\dimexpr\linewidth-9pt\relax}#2\end{minipage}}%
  \noindent\textcolor{#1}{\rule[-\dp0]{1.5pt}{\dimexpr\ht0+\dp0\relax}}\hspace{7pt}\usebox0\par}
\begin{figure*}[t]
\begin{minipage}[t]{\linewidth}
\footnotesize
{\scriptsize\bfseries\textsc{R2MED -- Medical-Sciences}}\\[2pt]
{\scriptsize\textcolor{black!55}{\textsc{query}}}\ \emph{Can pheromones or sun trigger/cause allergy?}\\[5pt]
\begin{tabular}{@{}lrrrr@{}}
\toprule
{\scriptsize\textsc{neurons}} & \parbox[b]{0.195\linewidth}{\raggedright\scriptsize\textcolor{black!55}{2746}\\\textbf{anaphylaxis}} & \parbox[b]{0.195\linewidth}{\raggedright\scriptsize\textcolor{black!55}{1573}\\\textbf{body temperature}} & \parbox[b]{0.195\linewidth}{\raggedright\scriptsize\textcolor{black!55}{1939}\\\textbf{human biology}} & \parbox[b]{0.195\linewidth}{\raggedright\scriptsize\textcolor{black!55}{411}\\\textbf{sun protection}} \\
\midrule
{\scriptsize\textsc{query}} & 0.395 & 0.183 & 0.239 & 0.190 \\
\midrule
\multicolumn{5}{@{}l@{}}{%
\cspassage{cslWon}{%
{\scriptsize\bfseries\textcolor{cslWon}{\textsc{Passage A -- Ranked 1 -- Label: Highly relevant}}}\newline
{\scriptsize\ Diseases \& Conditions Sun allergyOverview Sun allergy is a broad term. It describes several conditions that cause an itchy rash to form on the skin after being in sunlight or other sources of ultraviolet (UV) radiation. Polymorphous light eruption is the most common form of sun allergy. Some people have a hereditary type of sun allergy. Others develop symptoms only when triggered by another factor --- such as taking medication or touching certain plants. Other types of sun-related reactions occur for reasons that are unclear. Mild sun allergy may clear up without treatment. Severe rashes may be treated with steroid creams or pills. If you have severe sun allergy, you may need to take preventive steps. For example, wear clothing that shields you from the sun.}}} \\[3pt]
{\scriptsize\textsc{passage A}} & \textbf{\textcolor{cslWon}{0.380}} & 0.000 & 0.000 & \textbf{\textcolor{cslWon}{0.319}} \\
\midrule
\multicolumn{5}{@{}l@{}}{%
\cspassage{cslGradeZero}{%
{\scriptsize\bfseries\textcolor{cslGradeZero}{\textsc{Passage B -- Ranked 9 -- Label: Not relevant}}}\newline
{\scriptsize\ Different sources have different cut-offs for fever, hyperthermia and hyperpyrexia. vte In humans, hyperthermia is defined as a temperature greater than 37.5--38.3 $^{\circ}$C (99.5--100.9 $^{\circ}$F), depending on the reference used, that occurs without a change in the body's temperature set point.[3][10] The normal human body temperature can be as high as 37.7 $^{\circ}$C (99.9 $^{\circ}$F) in the late afternoon.[2] Hyperthermia requires an elevation from the temperature that would otherwise be expected. Such elevations range from mild to extreme; body temperatures above 40 $^{\circ}$C (104 $^{\circ}$F) can be life-threatening.}}} \\[3pt]
{\scriptsize\textsc{passage B}} & 0.097 & \textbf{\textcolor{cslGradeZero}{0.516}} & \textbf{\textcolor{cslGradeZero}{0.327}} & 0.000 \\
\bottomrule
\end{tabular}
\end{minipage}\\[14pt]
\begin{minipage}[t]{\linewidth}
\footnotesize
{\scriptsize\bfseries\textsc{NFCorpus}}\\[2pt]
{\scriptsize\textcolor{black!55}{\textsc{query}}}\ \emph{How to Help Prevent Abdominal Aortic Aneurysms}\\[5pt]
\begin{tabular}{@{}lrrrr@{}}
\toprule
{\scriptsize\textsc{neurons}} & \parbox[b]{0.195\linewidth}{\raggedright\scriptsize\textcolor{black!55}{1512}\\\textbf{venous thromboembolism prevention}} & \parbox[b]{0.195\linewidth}{\raggedright\scriptsize\textcolor{black!55}{1016}\\\textbf{abdominal wall}} & \parbox[b]{0.195\linewidth}{\raggedright\scriptsize\textcolor{black!55}{1163}\\\textbf{aneurysms}} & \parbox[b]{0.195\linewidth}{\raggedright\scriptsize\textcolor{black!55}{924}\\\textbf{aortic diseases}} \\
\midrule
{\scriptsize\textsc{query}} & 0.386 & 0.477 & 0.343 & 0.347 \\
\midrule
\multicolumn{5}{@{}l@{}}{%
\cspassage{cslWon}{%
{\scriptsize\bfseries\textcolor{cslWon}{\textsc{Passage A -- Ranked 1 -- Label: Highly relevant}}}\newline
{\scriptsize\emph{Analysis of risk factors for abdominal aortic aneurysm in a cohort of more than 3 million individuals.}}\newline
{\scriptsize\ BACKGROUND: Abdominal aortic aneurysm (AAA) disease is an insidious condition with an 85\% chance of death after rupture. Ultrasound screening can reduce mortality, but its use is advocated only for a limited subset of the population at risk. METHODS: We used data from a retrospective cohort of 3.1 million patients who completed a medical and lifestyle questionnaire and were evaluated by ultrasound imaging for the presence of AAA by Life Line Screening in 2003 to 2008. Risk factors associated with AAA were identified using multivariable logistic regression analysis. RESULTS: We observed a positive association with increasing years of smoking and cigarettes smoked and a negative association with smoking cessation. Excess weight was associated with increased risk, whereas exercise and consumption of nuts, vegetables, and fruits were associated with reduced risk\ldots{}}}} \\[3pt]
{\scriptsize\textsc{passage A}} & 0.000 & \textbf{\textcolor{cslWon}{0.261}} & \textbf{\textcolor{cslWon}{0.310}} & \textbf{\textcolor{cslWon}{0.290}} \\
\midrule
\multicolumn{5}{@{}l@{}}{%
\cspassage{cslGradeZero}{%
{\scriptsize\bfseries\textcolor{cslGradeZero}{\textsc{Passage B -- Ranked 2 -- Label: Not relevant}}}\newline
{\scriptsize\emph{Alzheimer's disease is incurable but preventable.}}\newline
{\scriptsize\ The dramatic rising incidence and costs of Alzheimer's disease (AD) require that research efforts and funding be primarily directed on either finding a cure or applying preventive measures to curb this disorder. A cure for AD appears unlikely when significant cognitive loss has occurred because the neuronal networks that controlled the perturbed cognitive abilities are either dead or irreversibly damaged and replacing them, even if it were technically possible, would not reconstruct the intellectual identity of the host. Prevention of risk factors to sporadic AD is a more realistic stratagem and treatment, when indicated, ideally should begin in cognitively intact individuals as part of a mass screening effort. Prevention of modifiable risk factors to AD is cost-effective because it reduces hospice or hospital stay, repeated doctor visits, and long-term care\ldots{}}}} \\[3pt]
{\scriptsize\textsc{passage B}} & \textbf{\textcolor{cslGradeZero}{0.331}} & 0.000 & 0.000 & 0.000 \\
\bottomrule
\end{tabular}
\end{minipage}
\caption{\textbf{Out-of-domain retrieval examples, complete passages.} Same CURE-trained SAE (BGE, $k{=}64$, $m{=}3072$) applied to two corpora it never saw, showing the four neurons that most separate each pair of passages. Each numeric row gives one activation per neuron, in the column order of the header above it. Stronger firing is bolded in the colour of the passage it favors. These same CURE-trained neurons account for \num{46}\% and \num{61}\% of the cosine similarity for the Medical-Sciences example, and \num{70}\% and \num{58}\% for the NFCorpus one, ranges comparable to the in-domain range.}
\label{fig:contrastive-full-ood}
\end{figure*}

\clearpage
\subsection{Dual topic query}
\label{twotopicquery}
We also illustrate how queries carrying two notions, here smoking and pregnancy, fire correctly the related neuron and how it can help us parse the SAE win with respect to the original model. 

We study the Medical-Sciences query \emph{``Is smoking during pregnancy harmful to child and mother?''}. The SAE ($k=64$, BGE, $m=3072$) raises NDCG@10 from \num{24.30} (original embeddings) to \num{36.33}. The original embeddings' top \num{5} contains no relevant passage. Under the SAE, two relevant passages move from ranks \num{7} and \num{10} into positions \num{3} and \num{4}. Three passages the original had in its top \num{5}, \texttt{Birth\_Defect\_31}, \texttt{Birth\_Defect\_43} and \texttt{Smoking\_During\_Pregnancy\_3} are pushed down. 

And we can see why:
\tableref{tab:maternalcontrib} shows all top 5 passages retrieved by the original embeddings, they all fire the \emph{smoking} feature hardest. It also shows how neuron 7, interpreted with TF-IDF terms as \texttt{women, maternal, psychological, woman, health}, contributes \num{0.068} and \num{0.045} on the two relevant passages the SAE moves up and exactly \num{0} on \texttt{Birth\_Defect\_43} and \texttt{Birth\_Defect\_15}, which are about \emph{paternal} smoking and instead fire neuron 869 \emph{parental influence}. Indeed, neuron 7 contributes \num{0.068} to score differences of \num{0.056} and \num{0.027} for \texttt{Smoking\_During\_Pregnancy\_1} over \texttt{Birth\_Defect\_43} and \texttt{Birth\_Defect\_15}, and \num{0.045} to differences of \num{0.040} and \num{0.012} for \texttt{Smoking\_During\_Pregnancy\_2}.

\begin{table*}[t]
\centering
\scriptsize
\floatconts
  {tab:maternalcontrib}
  {\caption{Union of top \num{5} passages retrieved for \emph{``Is smoking during pregnancy harmful to child and mother?''} by the original BGE embeddings and by the BGE-trained SAE ($k=64$, $m=3072$). \emph{rel.} marks the ground-truth relevant passages. Each neuron column is that
  feature's contribution $z_q[j]\,z_d[j]$ to the SAE cosine. Query activations: n2668 \num{0.593},  n1460 \num{0.279}, n7 \num{0.277}, n1976 \num{0.240}, n869 \num{0.188}.}}
  {%
  \setlength{\tabcolsep}{3pt}
  \setlength{\aboverulesep}{0pt}
  \setlength{\belowrulesep}{1pt}
  \renewcommand{\arraystretch}{1}
  \begin{tabular}{@{}l cc c rrrrr@{}}
  \toprule
   & \multicolumn{2}{c}{rank} & & n2668 & n1976 & n1460 & n7 & n869 \\
  \cmidrule(lr){2-3}
  Passage & orig. & SAE & rel. & smok. & pregn. & infant & matern. & patern. \\
  \midrule
  Smoking\_Pregnancy\_And\_Babies\_6 & 2 & 1 & & 0.284 & 0.099 & 0.097 & 0.082 & 0.000 \\
  \addlinespace[2pt]
  Smoking\_During\_Pregnancy\_5 & 12 & 2 & & 0.314 & 0.095 & 0.069 & 0.089 & 0.000 \\
  \addlinespace[2pt]
  Smoking\_During\_Pregnancy\_1 & 7 & 3 & $\checkmark$ & 0.250 & 0.073 & 0.112 & \textbf{0.068} & 0.000 \\
  \addlinespace[2pt]
  Smoking\_During\_Pregnancy\_2 & 10 & 4 & $\checkmark$ & 0.299 & 0.080 & 0.080 & \textbf{0.045} & 0.000 \\
  \addlinespace[2pt]
  Birth\_Defect\_15 \emph{(paternal)} & 4 & 5 & & 0.324 & 0.035 & 0.080 & \textbf{0.000} & 0.036 \\
  \midrule
  Birth\_Defect\_31 & 1 & 6 & & 0.177 & 0.065 & 0.129 & 0.067 & 0.023 \\
  \addlinespace[2pt]
  Birth\_Defect\_43 \emph{(paternal)} & 3 & 7 & & 0.261 & 0.048 & 0.073 & \textbf{0.000} & 0.039 \\
  \addlinespace[2pt]
  Smoking\_During\_Pregnancy\_3 & 5 & 8 & & 0.210 & 0.073 & 0.136 & 0.043 & 0.000 \\
  \bottomrule
  \end{tabular}
  }
\end{table*}

\section{Dead neuron analysis}
\label{sec:appdead}
\begin{table}[t!]
\centering
\small
\caption{Dead SAE latents, mean $\pm$ 95\% CI over 3 seeds. A latent is counted dead when it fires on neither the CURE training corpus nor the training queries. $k$: active SAE latents per embedding; $m$: SAE hidden width. \emph{+ k-aux}: trained with the auxiliary loss that makes the latents dead over a whole epoch reconstruct the residual of the main reconstruction. Configurations are selected on validation NDCG@10.}
\label{tab:deadlatents}
\begin{tabular}{l r r@{\;$\pm$\;}l r@{\;$\pm$\;}l}
\toprule
 & $k$ & \multicolumn{2}{c}{dead latents} & \multicolumn{2}{c}{\% of $m$} \\
\cmidrule(lr){3-4} \cmidrule(lr){5-6}
\multicolumn{6}{l}{\textbf{BGE} ($m$=3072)} \\
 & 64  &  167.3 &  39.6 &  5.45 & 1.29 \\
 & 128 &   36.3 &   2.9 &  1.18 & 0.09 \\
 & 256 &   29.7 &  20.7 &  0.97 & 0.67 \\
\midrule
\multicolumn{6}{l}{\textbf{Gemini} ($m$=9216)} \\
 & 64  & 2713.0 & 156.1 & 29.44 & 1.69 \\
 & 128 & 1296.0 & 147.9 & 14.06 & 1.61 \\
 & 256 &  374.3 &  71.8 &  4.06 & 0.78 \\
\midrule
\multicolumn{6}{l}{\textbf{Gemini + k-aux} ($m$=9216)} \\
 & 64  &   38.0 & 116.2 &  0.41 & 1.26 \\
 & 128 &  184.0 & 152.8 &  2.00 & 1.66 \\
\bottomrule
\end{tabular}
\end{table}

\tableref{tab:deadlatents} shows percentage of dead neurons (neurons that never fire on either the training corpus or the training queries) on BGE and Gemini embedders. To try to decrease \% of dead neurons on Gemini, we added the auxiliary loss to prevent dead neurons via reconstruction of the residuals \citep{oneill2024disentangling, gao2024scaling} and swept over $k=64,128$ for Gemini. This cuts dead neurons from $2713 \pm 156$ to $38 \pm 116$ at $k=64$ and from $1296 \pm 148$ to $184 \pm 153$ at $k=128$. Retrieval performance however is either on par or lower. Indeed, in domain, CURE is on par at both $k$. Out of domain, \num{3} of the \num{10} comparisons (\num{5} datasets $\times$ \num{2} values of $k$) are lower: Medical Sciences at $k=64$ ($39.57 \to 33.12$), NFCorpus and MIRIAD at $k=128$ ($35.19 \to 33.08$ and $79.07 \to 76.78$), and the remaining \num{7} are on par. Thus, we leave this exploration of trade-off between latent liveness and retrieval quality to future work.

\end{document}